\documentstyle[aas2pp4]{article}

\begin{document}
\title{Early Star Formation and Chemical Evolution in Proto-Galactic 
Clouds}

\author{Lamya Saleh}
\affil{Department of Physics \& Astronomy, Michigan State University, E.
Lansing, MI  48824\\email:  saleh@pa.msu.edu}
\author{Timothy C. Beers}
\affil{Department of Physics \& Astronomy and JINA: Joint Institute for Nuclear
Astrophysics, Michigan State University, E.
Lansing, MI  48824\\email:  beers@pa.msu.edu}
\author{Grant J. Mathews}
\affil{Center for Astrophysics, 
Department of Physics,
University of Notre Dame,
Notre Dame, IN 46556\\email: gmathews@nd.edu}

\date{\today}
\begin{abstract}

We present numerical simulations to describe the evolution of pre-Galactic
clouds in a model which is motivated by cold dark matter simulations of
hierarchical galaxy formation. We adopt a SN-induced star-formation mechanism
within a model that follows the evolution of chemical enrichment and energy
input to the clouds by Type II and Type Ia supernovae. We utilize
metallicity-dependent yields for all elements at all times, and include effects
of finite stellar lifetimes. We derive the metallicity distribution functions
for stars in the clouds, their age-metallicity relation, and relative elemental
abundances for a number of alpha- and Fe-group elements. The stability of these
clouds against destruction is discussed, and results are compared for different
initial mass functions. We find that the dispersion of the metallicity
distribution function observed in the outer halo is naturally reproduced by
contributions from many clouds with different initial conditions. The scatter in
metallicity as a function of age for these stars is very large, implying that no
age-metallicity relation exists in the early stages of galaxy formation. Clouds
with initial masses $^>_\sim$ presently observed globular clusters are found
to survive the first 0.1 Gyr from the onset of star formation,
suggesting that such systems may have contributed to the formation of
the first stars, and could have been self-enriched. More massive clouds are
only stable when one assumes an initial mass function that is not biased towards
massive stars, indicating that even if the first stars were formed according to
a top-heavy mass function, subsequent star formation was likely to have
proceeded with a present-day mass function, or happened in an episodic manner.
The predicted relative abundances of some alpha- and Fe-group elements show reasonable
agreement with the observed values down to metallicities below [Fe/H] $\sim -4$
when the iron yields are reduced relative to stellar models.
The observed scatter is  also reproduced for most elements including the observed
bifurcation in [$\alpha$/Fe] for stars with low [Fe/H]. However,
the predicted dispersion may be too large for some  elements (particularly alpha elements)
unless a limited range of progenitor masses contributing to
the abundances of these elements is assumed. The contributions to the
abundances from supernovae with different progenitor masses and metallicity are
discussed. The results suggest that the low-mass end of SNeII  was
probably absent at the very lowest metallicities, and that the upper mass limit
for the first stars that contributed to nucleosynthesis may be $^<_\sim 40$
M$_\odot$. 

\end{abstract}

\section{Introduction}

The oldest and most metal-poor stars observed in the Galaxy are mainly
identified as members of the halo population, and are believed to have formed at
the earliest times in the history of the proto-Galaxy. These stars are expected
to have contributed to and influenced the nature of all of the various
populations observed today, through their formation processes and the dynamical
interactions that followed. Thus, their chemical composition and kinematic
properties can provide clues to the early epochs of Galaxy formation.
Accordingly, many studies have been performed to identify trends in the chemical
composition of halo stars and their connection to spatial and kinematic
properties (Mathews, Bazan \& Cowan 1992; McWilliam 1995; Ryan, Norris, \& Beers
1996; King 1997; Carney et al. 1997; Preston \& Sneden 2000; Kinman et al.~2000;
Argast, et al.~2000; 2002; Carretta et al. 2002; 
Cohen et al.~2002; 2004; Arnone 2004; Cayrel, et al. 2004). 

A dispersion in the abundances of heavy elements, varying from less than an
order of magnitude for alpha and iron-group elements, to more than two orders of
magnitude for neutron-capture elements has been detected at very low
metallicities. This has led to the suggestion that the lowest-metallicity stars
were the result of mixing of the ejecta from single, or at most a few,
supernovae, with a limited amount of gas in the parent clouds (Audouze \& Silk
1995; Ryan et al. 1996; McWilliam 1997; 1998 Norris et al. 2000).
However, the fact that the dispersion in Mg/Fe is so small
(Arnone et al. 2004)
may indicate that more efficient mixing may have occurred for this element
and perhaps other alpha elements as well
(Carretta et al. 2002; Cayrel et al. 2004; Cohen et al. 2004).

In addition, several studies (Lance 1988; Preston, Shectman, \& Beers 1991;
Preston, Beers, \& Shectman 1994; Majewski 1992; Kinman et al.~1994; Carney et
al. 1996; Sommer-Larsen et al. 1997; Lee,\& Carney 1999; Chiba \& Beers 2000;
Chiba \& Beers 2001) have suggested a possible duality of the halo population.
The observed kinematics and ages of halo field stars suggests a hybrid formation
process in which the inner halo may have formed by a coherent contraction in the
manner originally suggested by Eggen, Lynden-Bell, \& Sandage (1962) [henceforth
ELS] (see also Carney et al. 1996; Sommer-Larsen et al. 1997; Chiba \& Beers
2000) while the outer halo was formed by accretion of metal- poor fragments from
nearby systems (e.g., Chiba \& Beers 2000, and references therein).

The most favored current theory for galaxy formation is based on a
hierarchical clustering scenario in which the Galaxy is assembled from cold
dark matter sub-galactic halos (e.g., Peacock 1999). These proto-galactic clumps are
believed to have originated from density fluctuations in the early universe
that accreted primordial gas from their surroundings. Merging processes would
lead to larger structures, and ultimately, the Galaxy.  Cold dark matter (CDM)
models are consistent with the observed kinematics of the halo of the Galaxy
(Chiba \& Beers 2000) and with the suggested hybrid formation scenario, but
still require intensive additional observations and testing to quantify the
parameters required to produce these trends in detail.
  
Chemical evolution models that have treated these early stages still lack a
complete picture that includes all chemical and dynamical effects. They are not
yet capable of providing a comprehensive picture that reproduces the scatter in
observed abundances at low metallicities. Some models in the literature (Malinie
et al.~1993; Ishimaru \& Wanajo 1999; McWilliam \& Searle 1999; Raiteri et al.
1999; Tsujimoto et al.~1999; Argast et. al.~2000; 2002; Travaglio et al.~2000; Oey
2003) have been able to reproduce, at least partially, the inhomogeneities
observed for elemental abundances through a stochastic evolution scenario. These
are, however, often limited for example to the study of only a few
neutron-capture elements, and/or metallicity independent yields, and/or
instantaneous recycling. In the present paper we
incorporate metallicity dependent stellar ejecta and finite stellar
lifetimes.
  
The enrichment processes that took place in these first structures were
influenced by the initial conditions in each cloud, e.g. the initial cloud
mass. While smaller clouds are susceptible to destruction by tidal forces and
energetic feedback and may experience shorter episodes of chemical evolution,
larger structures may have continued to enrich the interstellar media (ISM) for
longer periods of time and survived the merging process, contributing
eventually to the formation of the disk.

In this work we present a simple chemical-evolution model in which we 
evolve pre-Galactic clouds in a scenario that incorporates the
main features of hierarchical galaxy formation
as suggested by cold dark matter simulations. 
Our model is based upon several assumptions.  These are:
1)   In the first zero-metallicity clouds, the first-generation stars are born
as population III (Pop III), explode, and the clouds that survive are enriched by their
expelled yields; 
2)  Star formation is induced by supernovae, mainly
Type II (SNeII);
3) Each individual
supernova is taken to form a shell, which expands with an average
velocity and and triggers star formation
with an average efficiency $\epsilon$ over a timescale $\Delta t$;
4)  The shells' mass is the sum of the whole SNII gas and the
swept-up surroundings, and the metallicity is given by a complete
mixture of both metal contents; 
5)   The SN II yields are taken from the literature as metal dependent; and  
6)  The stellar initial mass function (IMF) 
according to various conditions within the cloud.

These assumptions, though adequate for the phenomenology of interest here,
nevertheless warrant some caveats. For example, under the first assumption, one
should in principle take care of details in shell fragmentation theory of shell
enrichment. In the present work this is accomplished with effective star
formation rates averaged over the shell.

Although assumption 2 is an appropriate scenario for enhanced star formation, it
neglects some hydrodynamic negative self-regulation. For example, the
hydrodynamic expansion of a SN bubble depends on the external density. In the
hydrodynamic Sedov phase (Sedov 1946; Ostriker \& McKee 1988;  Chioffi, McKee \&
Bertschinger 1988) energy is lost due to expansion because the expansion velocity
decreases with time, and also depends on density as $\rho^{-1/5}$. When the
ambient density is larger at an earlier times, both radius and velocity are
less, so the sweep-up of surrounding gas could be smaller than assumed here.
Furthermore, if subsequent stars are formed in associations, SNII bubbles of
massive stars would be expected to overlap and to form a so-called super bubble.
This means that the shells' total surface is smaller, and the amount of swept-up
gas is reduced.

In our model all of these details are absorbed into simple average properties of
the clouds and shells. Nevertheless, these assumptions are adequate to describe
the bulk of the observations, as we shall see. In particular, we incorporate
the stochastic nature by which the initial conditions including total mass, and
metallicity must have been distributed among the different clouds according to a
hierarchical clustering scenario. This random nature is further extended during
the early stages of evolution by the dependence of the chemical yields produced
in SN events on the initial mass and metallicity of their progenitors.
Therefore, we consider clouds of different initial masses and form stars
according to four different IMFs. We also consider {\it metallicity-dependent}
chemical yields of a number of iron group and alpha elements allowing for the
products of SN events to change with time. 

Assumption 4 is another very crucial assumption, since it strongly enters the
metallicity results. The interstellar medium of galaxies includes a hot
metal-rich gas phase that also carries these elements and could evaporate them
from the clouds. Hence, this assumption is relevant for element abundances and
the timescale of their enrichment. Indeed, dwarf galaxies may require
inefficient mixing of supernova ejecta with the ISM to achieve observed low
oxygen abundances and other low effective yields. We note, however, that
although the full yields of SNeII may not be instantaneously incorporated and
homogeneously mixed into the cool gas, in our model this loss of SN mixing
efficiency can be compensated for by an increase in the star formation efficiency.
Hence, for our purposes an efficient mixing of the supernova ejecta is an
adequate working hypothesis. We return to this point below.

The specific form for SN-induced star formation in our model is based upon that
proposed by Tsujimoto et. al (1999). That is, star formation is triggered by SN
events, and the material from which the next generation of stars forms is the
result of complete mixing of the SN ejecta with the gas swept up from the ISM by
the explosion. This produces low-metallicity stars with elemental abundances
resembling the ejecta of high-mass SN progenitors, as suggested by the data.

We show that this simple stochastic model is capable of reproducing the observed
scatter in abundances, relative to iron, at low metallicity, as well as the
observed shift in trends at higher metallicities after the chemical products of
stars mix with the ISM. The metallicity-dependent chemical yields of Woosley \&
Weaver (1995, hereafter WW95) used in this model, have been used previously in
detail by Timmes et al. (1995), who showed general agreement with the observed
relative abundances in the Solar neighborhood for most elements. They did not,
however, attempt to apply these chemical yields to a stochastic, CDM-motivated
model for the halo, and they only did their comparison for [Fe/H] $\ge -3$.
  
In the present work we test these yields with our model for several alpha and
several iron-group elements down to metallicities below [Fe/H] $\sim -4.0$, near
the present observational limit of metallicity in the Galaxy.

\section{The Model}

The main purpose of the present study is to explore the viability of a simple
schematic hierarchical clustering paradigm for galaxy formation. In this
picture, primordial fluctuations lead to the gravitational collapse of
subsystems forming sub-galactic halos comprised of both baryons and dark matter.
Such subsystems then merge and accrete diffuse matter to form proto-galactic
halos. We follow the idealized chemical evolution of the baryon component of
individual clouds, assuming all stars formed initially in proto-Galactic clouds
via SN-induced star formation (e.g., Tsujimoto et al. 1999). We utilize
metallicity-dependent chemical yields and follow the evolution of several alpha-
and Fe-group elements. These are compared to the observed abundances of
individual stars in the halo and thick disk of the Galaxy.

Our model consists of evolving individual clouds in the mass range of $10^5$ to
$10^8$ M$\odot$. We follow the chemical evolution of each cloud as a one-zone
closed box (Tinsley 1980). The evolution of gas and stars is then explicitly
evolved in time. We assume that all halo field stars were formed originally in
these proto-Galactic clouds and were later dispersed by tidal forces. By varying
the total masses and the star formation histories of these clouds, we attempt to
account for the trends and scatter in elemental abundances observed in
metal-poor stars of the halo and thick-disk populations.

Simulations show that as the baryonic component of these clouds cools (mostly
via molecular hydrogen) they gravitate to the center of the dark-matter
potential to form central cores (Abel et al.~1999; Norman et al. 2000). These
cores are believed to form stars at their centers, the exact nature of which is
still uncertain (Silk 1983; Carr et al. 1984). We assume that each cloud starts
with primordial material, and ultimately forms one massive star at its center.
This Pop-III star later explodes as a SN event, triggering the
formation of higher metallicity (population II) stars in its high-density shell.
Subsequent star formation progresses in the shells of later SN events.

Star formation has been linked to the dynamical evolution of molecular clouds in
many studies. Observations of nearby molecular clouds suggest that star
formation is triggered by expanding shells driven by stellar winds and
repeated SN events in galaxies (Yamaguchi et al. 2001; Hartmann 2002). When
these shells become unstable and fragment, they may form molecular clouds and
stars (Ehlerova et al. 1997; Blondin et al. 1998). The conditions under which
star formation is triggered due to the gravitational collapse of expanding
shells has been investigated by Elmegreen et al. (2002) with a 3-D numerical
simulation of the conditions in the ISM as the expanding shell injects its
energy. 

In all of these studies, stars are assumed to form out of the gas in the ISM 
and reflect it¼s metal content at the time of their formation. Although this
treatment of the star formation process in galactic chemical evolution models has been
very successful in reproducing the general features of  the chemical
compositions of stars and HII regions in the solar neighborhood (Matteucci \&
Greggio 1986; Yoshii, Tsujimoto, \& Nomoto 1996), it cannot be applied to the
early stages of the galaxy¼s evolution (Shigeyama \& Tsujimoto 1998), as
it fails to reproduce the observed abundance patterns in extremely
metal-deficient stars (Mcwilliam et al. 1995; Ryan et al. 1996). Therefore, an
alternative scenario was proposed by Tsujimoto et al. (1999) in which stars are
born in the dense shells formed by the sweeping up of ISM during a SN
explosion.  Ryu \& Vishniac (1987) suggest that such dense shells which form
behind the radiative shock front are  dynamically overstable and are broken
into a few thousand fragments in which self gravity is unimportant. Nakano
(1998) suggests that some of these fragments will dynamically contract and
eventually form stars.  Such a SN-induced star formation scenario has been shown
to fit the observed relative abundances at early times better than models where
SN ejecta mix completely with all the gas in the ISM (Nakamura et al.  1999). 
We adopt this scenario with a new set of equations consistent with our use of
metallicity-dependent yields.  The metallicity of each shell is calculated at
the time of its formation and stored for use in subsequent generations.

This scenario leads to a difference between the abundances observed in stars and
those of the ISM at the time they were formed. This differs from most Galactic
chemical evolution models, in which the stellar and gas abundances are
constrained to be identical. The material from which the next generation of
stars forms, following a given SN explosion, is the result of the complete
mixing of the SN ejecta with the gas swept up from the ISM by the explosion.
This star formation scenario will produce low-metallicity stars with elemental
abundances resembling that of a single high-mass supernova. Our model follows
the evolution of gas in the clouds and the evolution of individual elements in
the ISM. We also follow the chemical evolution of stars and derive the
anticipated present-day metallicity distribution function for halo stars.
  
\subsection{The Calculation}

We performed numerical solutions of the chemical-evolution equations for each
cloud, deducing the evolution of gas mass, $M_g$, and 
mass of individual elements, $M_i$, in finite steps of time. At
each time step, the equations are integrated over a grid of stellar masses,
ranging from 0.09 to 40 M$_\odot$. In the context of this model, we desire a time
step, $\Delta t$, longer than the time required for a SN shell to form and
disperse, but short enough to be sensitive to the short lifetimes of massive
stars. For example, the expansion time of a SN event is of the order of 0.01 Myr
(Binney \& Tremaine 1987), while the lifetime of the most massive stars included
in our model is $\sim 6$ Myrs. Since the sensitivity to the lifetimes is very
crucial at the earliest times, the time step was taken to be $\sim 1$ Myr up to
the point when the least-massive SNe would have contributed ($\sim$20 Myrs).
Thereafter, $\Delta t$ was given a larger value of 25 Myrs.

We take into consideration the main sequence lifetimes of these stars, relaxing
the instantaneous recycling approximation. This is of fundamental importance
when dealing with timescales as short as the halo formation time, and when
treating intermediate-mass elements, which receive significant contributions
from progenitor stars with lifetimes comparable to the halo formation time. We
adopt the mass-dependent lifetimes used by Timmes et al.~(1995), i.e.~the
lifetimes of Schaller et al. (1992) were used for stars less massive than 11
M$_\odot$, while   the lifetimes given by the stellar
evolution calculations of Weaver \& Woosley (1994) were adopted
for more massive stars

For a given cloud, when the first shell forms at $t = 0$, the rate at which
mass goes into stars in the cloud at that time is 
\begin{equation}
\psi(t=0) = \epsilon M_{sh}(m,0)/\Delta t~~,
\end{equation}
 where $\epsilon$ is the star formation efficiency which gives
the mass fraction of the shell that is used up in 
forming stars.  It is treated as a free parameter in our model. 
Changing this parameter can
only have an effect on the time scales, but will not alter the overall trends
calculated for the elemental abundances or the MDFs. The $\epsilon$ parameter
was adjusted in this model to reproduce the shift in slope of the elemental
abundance ratios for three iron-group elements below [Fe/H] $\sim -2.4$ observed
by McWilliam et al. (1995). It was also made to be consistent with the
observation that the age-metallicity distribution changes for extremely metal-
poor stars with [Fe/H] $< -2.4$. $M_{sh}(m,t)$ is the mass of the shell formed
in a SN explosion at time $t$, with the progenitor mass being $m$, and is given
by :
\begin{equation}
M_{sh}(m,t) = E_j(m,Z) + M_{sw}~~,
\label{msheq}
\end{equation}  
where $M_{sw}$ is the mass of the interstellar gas swept up by the
expansion.  Since the explosion energy of a core-collapse supernova depends only
weakly on the progenitor mass (WW95; Thielemann et al. 1996), this quantity is
taken to be constant with a value of $5\times 10^4$ M$_\odot$ (Ryan et al. 1996;
Shigeyama \& Tsujimoto 1998; Tsujimoto et al. 1999). $E_j(m, Z)$ is the mass of
all the ejected material from the SN with a progenitor mass $m$ and metallicity
$Z$. [Note that if not all of the supernova ejecta is mixed into the ISM, then
this can be compensated in out model by the star-formation parameter $\epsilon$.]
For the first SN event the metallicity $Z$ is zero for a population III
star, and $ m$ is $m_1$. For later generations, the metallicity of a star is
calculated from the metallicity of the shell from which it was formed,
$Z_{sh}(m,t)$. This is calculated as a function of time, and is given by:

\begin{equation}
Z_{sh}(m,t) = 1 - [x^H_{sh}(m,t) +  x^{He}_{sh}(m,t)]~~.
\end{equation}

\noindent Here, $x^ H_{sh}$ and $x^{He}_{sh}$ are the fractions of H and He in the
shell, respectively, and are given by:

\begin{equation}
x^H_{sh}(m,t) = [y^H(m,t)  +  x^H_{gas}(t)M_{sw}]/M_{sh}(m,t)~~,
\end{equation}

\begin{equation}
x^{He}_{sh}(m,t) = [y^{He}(m,t)  +  x^{He}_{gas}(t)M_{sw}]/M_{sh}(m,t)~~,
\end{equation}

\noindent where  $y^H$ and $y^{He}$ are the mass of ejected H and He, respectively,
from the explosion. We will refer to these quantities as the ``yields''. They
are metallicity dependent and therefore are functions of time. The quantities
$x^H_{gas}$ and $x^{He}_{gas}$ are the fractions of H and He in the interstellar
gas at the time of formation of the shell $t$, respectively. This fraction is
given for any element $x^i_{gas}$ by:

\begin{equation}
x^i_{gas}(t) = M_i(t)/M_g(t)~~.
\end{equation}

For the first explosion, $y_i(m_1,t_1)$ is the yield of elements $i$ from stars
of mass $m_1$ and metallicity zero. The subsequent star formation
rate (SFR), $\psi(t)$, is derived from a sum over all shells that form at time $t$. It thus
depends on the SFR at the time the progenitor star formed, $\psi(t-\tau_m)$,
where $\tau_m$ is the lifetime of the progenitor.

\begin{eqnarray}
\psi(t>0) &= &\int_{max(m_t,10)}^{m_u} \epsilon M_{sh}(m_i,0)\biggl[{\phi(m)\over m}\biggr] 
\nonumber \\
& \times &
\psi(t- \tau_m) dm
\label{psitgt}
\end{eqnarray}

\noindent where $\phi(m)$ is the IMF, normalized such that

\begin{equation}
\int_{m_l}^{m_u} \phi(m) dm = 1~.
\end{equation} 

\noindent The quantities $m_u$ and $m_l$ denote, respectively, the upper and
lower mass limits for the IMF. The lower mass limit for stars that produce SN
events of type-II will be taken as 10 M$_\odot$. The quantity $m_t$ is the mass
of a star with lifetime equal to $t$, measured from the time of formation of the
first shell. Since the ejected mass from a SN changes with the metallicity of
the progenitor, the quantity $E_j(m,Z_{sh})$ in Eq. (\ref{msheq}), must be
replaced by an average over all stars of mass m and different values of
$Z_{sh}$, which explode at a given time $t$:

\begin{eqnarray}
&&\langle M_{ej}(m,t) \rangle  =\int_{max\{m(t-\tau_m),10\}}^{m_u} dm' 
\nonumber \\
&&\times 
\biggl[(\phi(m')/m') m_{ej}(m,Z_{sh}(m',t-\tau_m)\biggr]~,
\end{eqnarray}

\noindent Here, $m'$ is the mass of the progenitor which produces the shell 
from which $m$ is formed, and $m_{ej}(m, Z_{sh}(m',t-\tau_m))$ is the mass
ejected from the star with mass $m$ and metallicity $Z_{sh}(m',t-_m)$. The
quantity $(t - \tau_m)$ denotes both the time of formation of a star of mass $m$
and the death of the star $m'$.  

The same applies for the yields of individual elements, $y_i$
\begin{eqnarray}
\langle y_i(m,t) \rangle &=& \int_{max\{m(t-\tau_m),10\}}^{m_u} dm'u
\biggl[\phi(m')/m'\biggr]
\nonumber \\
 &\times &y_i^{ej}(m,Z_{sh}(m',t-\tau_m))~~.
\end{eqnarray}
and $y_i^{ej}(m, Z_{sh}(m',t-\tau_m))$  is the mass of the element
$i$ ejected from the star with progenitor 
mass $m$ and metallicity $Z_{sh}(m',t-\tau_m)$.

The change in gas mass with time is then given by:
\begin{eqnarray}
&&{dM_g \over dt} =  -\psi(t) + 
\int_{max(m_t,m_l)}^{m_u} dm [\phi(m)/m] 
\nonumber \\
&& 
\times M_{ej}(m,t) \psi(t-\tau_m)~~. 
\end{eqnarray}

\noindent 
The first term in this equation is the SFR, equal to the amount of
gas going into forming stars at time $t$.  It is derived from Eq.     
(\ref{psitgt}).  The second term accounts for the enrichment process by all stars
whose life ends at time  $t$, including the whole possible range from
$m_t$ to $m_u$.  The change in the mass of element $i$ in the gas
is given by:

\begin{eqnarray}
&&{dM_i \over dt}  = - \int_{max(m_t,10)}^{m_u} dm [\phi(m)/m] 
\nonumber \\
&&\times  \epsilon M_{sh}(m,t) x_i(m,t)  \psi(t-\tau_m) \nonumber \\
&+&\int_{max(m_t,m_l)}^{m_u} dm [\phi(m)/m]
\nonumber \\
&&\times  y_i(m,t)  \psi(t-\tau_m) ~~.  
\end{eqnarray}                                                                                                

\noindent The first term in this equation gives the rate at which element
$i$ is incorporated into stars at time $t$, whereas the second term represents
the rate at which it is being added to the ISM by stars. By substituting Eqs.
(7)-(9) into equations (10) and (11), this set of integro-differential equations
is solved numerically. An auto-regressive computation of the star formation rate
and the metallicity of each shell that forms was employed, and previous values
were substituted in equations (7) - (11) at each time step. At every time step,
the metallicities of all shells formed are calculated and stored for use at
later times. Each shell is recorded and the age-metallicity relation is
constructed. The evolution of the metal content in both the gas and and the
stars are followed as a function of time.

\subsection{Metallicity Dependent Chemical Yields}

\subsubsection{SNII Yields}

In this model we have utilized the stellar yields calculated by WW95 for high-
mass stars. These are more-or-less consistent with the yields of Portinari,
Chiosi \& Bressan (1998). Our study is therefore complementary to the models of
Argast et al. (2000; 2002) who made a similar stochastic study, but relied upon the
yields of Thielemann, Nomoto, \& Hashimoto (1996) and Nomoto et al. (1997). We
use finely spaced mass and metallicity grids in our simulation, and linearly
interpolate the yields in both dimensions. The yields of every SN event are
mass- and metallicity-dependent, and the metallicities are calculated for all
stars at the time of their formation and stored for use at later times. WW95
included stars with progenitor masses ranging from 11 to 40 M$_\odot$, and
metallicities from zero to solar, and to calculate yields for elements from H to
Zn. This permits better limits to be set on Galactic chemical evolution models,
and enables comparisons between the results of simulations and the empirical
data from high-resolution spectral measurements in stars.

The uncertainties associated with the use of such supernova yields have been
critically summarized in Argast et al. (2000; 2002). In particular, the synthesized
yields can be sensitive to various aspects of both the stellar nucleosynthesis
models and the supernova explosion mechanism and energy released. As a result
one needs to be cautious in inferring quantitative conclusions regarding
detailed elemental abundances. Nevertheless, qualitative trends in elemental
ratios can be studied. By exploring the trends based upon independent models
utilizing different independently derived yields it is at least hoped that the
present study will help to clarify the underlying issues in early Galactic
chemical evolution. In this sense it is important to compare
the present study with that of Argast et
al. (2000).

It is useful to consider some of the differences between our yields and those
employed by Argast et al. (2000). While the yields presented by WW95 cover stars
up to 40 M$_\odot$, other studies, such as Thielemann, Nomoto \& Hashimoto
(1996) obtain yields for stars up to 70 M$_\odot$. Although the fate of stars
more massive than 40 M$_\odot$ is still uncertain, recent studies have suggested
that metal-free stars with masses between 35-100 M$_\odot$ collapse into black
holes, while stars with masses between 10-35 M$_\odot$ can explode as SNe-II, 
and contribute to the enrichment process ( Heger, Woosley, \& Waters 2000;
Heger et al. 2002). In addition, including such massive
stars will have very little effect on the results of a chemical-evolution
calculation when adopting a Salpeter IMF, since very few stars will form that
are more massive than 40 M$_\odot$. The metallicity dependence of the yields of
some elements, such as oxygen, is very strong for such massive stars.
This dependence will influence greatly their evolution at early times.
Since oxygen is not included in our study, we believe that the inclusion of
these massive first stars is not crucial here. Therefore, the stellar mass range
encompassed by the calculations of WW95 is reasonable. This is also consistent
with the findings of Samland (1998).

In contrast to the stellar yields calculated by Thielemann et al. (1996), in
which they use a constant solar progenitor metallicity, the Woosley \& Weaver
yields have the advantage of covering the whole metallicity range from zero to
solar for the progenitor stars.  Although their predicted yields of heavy
elements are not affected greatly by small changes in the initial metallicity
of the progenitor (for $Z/ Z_\odot > 0$), there are large differences between the
predicted yields for stars with metallicity $Z/ Z_\odot > 0$ and those for
stars with $Z/ Z_\odot= 0$.  This has a significant effect on the metallicity
distribution of extremely metal-poor stars, and on the stars of the next
generation that will form in their shells.

WW95 also emphasized that, in addition to the initial mass and metallicity, the
energy of the explosion is an important parameter in determining the yields from
a SN event, The energy of the explosion determines the mass cutoff between the
part of the synthesized elements that will be ejected and elements in the deeper
layers that will fall down onto the core after the explosion. They distinguished
different models A, B, and C, for stars more massive than 25 M$_\odot$. These
correspond to three different values for the initial kinetic energy of the
``piston'' (a theoretical construct used to simulate a real explosion). The
larger this energy, the more heavy elements escape from the explosion. We adopt
model B, an intermediate value for the energy.

\subsubsection{Intermediate-Mass Yields}

For intermediate stellar masses we adopt the standard nucleosynthesis yields of
Renzini \& Voli (1981). Although other tracks are available (e.g. Portinari et
al. 1998), the Renzini and Voli (1981) tracks are sufficient for the present
study, which is almost unaffected by the yields of intermediate mass stars other
than through the overall evolution of metallicity. The Renzini \& Voli (1981)
yields are also the same as those adopted in earlier works (e.g. Timmes et al.
1995) and so are useful for comparison of the present stochastic model with the
results of that earlier continuous-star-formation model.

The Renzini \& Voli (1981) models included stars in the mass range of $1 < m < 8
M_\odot$. They calculated the yields for two different metallicities, $Z =
0.004$, and $Z = 0.02$. In the current model, we interpolated between these two
values and used $Z = .004$ yields for lower metallicities. Intermediate-mass
stars mainly eject H, He, C, and N into the ISM at the end of their lives. Thus
they mainly only affect the metal/H ratio. 

\subsubsection{SNIa Yields}

Intermediate-mass stars mostly end their lives as carbon-oxygen rich
white dwarfs.  If they are members of a binary system these dwarfs may accrete
mass from the remaining member to the point where their mass exceeds the
Chandrasekhar limit.  At this point, the star becomes unstable, causing a
thermonuclear Type-Ia SN event.  For these events we use the yields calculated by
Thielemann et al. (1986).    

The appearance of the Fe-rich material contributed by Type-Ia SN is delayed by
the time required by a white dwarf to evolve to an explosion. This delay time has
been estimated to be of the order of 1 Gyr (Greggio \& Renzini 1983; Matteucci
\& Greggio 1986; Smecker \& Wyse 1991; Mathews et al.~1992; Yungelson \& Livio
1998; Tantalo \& Chiosi 2002). Therefore, these events will only have a
significant effect on clouds that evolve and self-enrich on a timescale of more
than 1 Gyr. 

Disrupted clusters disperse and permeate the halo. The dispersed
remnants can produce isolated SNeIa but should not significantly affect the
cloud abundances of interest in the present study. Only SNIa events which occur
within the clouds affect the present study, which can be the case for clouds
that are sufficiently massive to retain their gas. Less massive structures are
expected to be destroyed due to the energy input from SN of Type-II events during
the first Gyr of the onset of star formation. Stars less massive than 1
M$_\odot$ serve in this model only as reservoirs of gas mass, since they do not
evolve significantly during the considered time.

\subsection{The IMF}

In addition to the type of mixing processes following the explosion, the
enrichment of the halo depends on the type and number of SN events that took
place. Therefore, the central role that the IMF plays at these early times
cannot be ignored. For a SN-induced star formation model, the SFR depends on the
rate of death of massive stars at a given time. In other words, the SFR today
depends on the SFR at previous times. Therefore, the IMF of the first stars
(perhaps more properly referred to as the First Mass Function, FMF) will affect
greatly the subsequent evolution of these clouds. 
  
In this regard it is of interest that several recent studies have suggested that
Pop-III stars where probably very massive (Elmegreen 2000a; Bromm et al.~2001b;
Bromm et al. 2002; Abel et al. 2002; Kroupa 2002; Nakamura \& Umemura 2002), or
may have had a very massive component (Nakamura \& Umemura 2002) compared to
modern stellar populations. 

Although the shape of the IMF that dominated in
primordial conditions is still uncertain, there is compelling circumstantial
evidence pointing toward an early IMF biased towards more massive stars. This
evidence includes the paucity of metal-poor stars in the Solar neighborhood
[i.e.~the G-dwarf problem (Schmidt 1963; Larson 1998)], observations indicating
a rapid early enrichment of the ISM by heavy elements produced in the explosions
of high-mass stars (Mushotzky et al. 1996; Loewenstein \& Mushotzky 1996),
followed by slower enrichment at later times, represented by a flat
age-metallicity relation in the solar neighborhood (Larson 1986). Further
evidence is found in the high abundances of heavy elements in the hot gas
trapped in the potential wells of rich clusters of galaxies. The heavy-element
abundance is larger than what would be predicted by a standard present-day IMF
(Zepf \& Silk 1996 and references therein; Larson 1998). 

Moreover, several
studies of the central regions of star burst galaxies have suggested that the IMF is
strongly biased towards high-mass stars (Doane \& Mathews 1993; Rieke et al.
1993; Doyon, Joseph, \& Wright 1994). Observations of nearby ellipticals show
that [Mg/Fe] increases above the solar value with increasing galaxy luminosity
(Worthey, Faber \& Gonzalez 1992; Davies, Sadler , \& Peletier 1993) , and of
distant galaxies with unusually red colors or strong 4000 A breaks (Charlot et
al. 1993). These results are suggestive of an IMF that was once weighted toward
massive stars, perhaps during the early merger that was the hallmark of a
forming elliptical.
 
A top-heavy IMF during the initial stage of elliptical galaxy formation has 
also been proposed as a way to account for the large iron masses in galaxy 
clusters (Elbaz, Arnaud, \& Vangioni-Flam 1995). Recent observations of the 
elemental abundances of the intracluster gas point to Type II SN as the source
of enrichment, providing strong support for models with a top-heavy IMF 
during the formation of elliptical galaxies (Loewenstein \& Mushotzky 1996).

In addition, the elemental abundances observed in halo stars suggest that they
were made out of gas enriched only by SNeII      (Nissen et al. 1994; Ryan et
al. 1996; Tsujimoto et al.~1999; Norris et al.~2000). Finally, recent
theoretical calculations (Nakamura \& Umemura 2002) of star formation in
extremely metal-deficient clouds suggests that, in such conditions, the mass
function is peaked at the high mass end ~ (10-100 M$_\odot$) and is deficient in
sub-solar mass stars.

The Salpeter power-law IMF represents the stellar distribution in the
solar neighborhood very well down to 1 M$_\odot$.  At lower masses the form of
the function is not clear due to the difficulty of obtaining a mass-luminosity
relation for such faint stars.  However, it is believed that it must decline
rapidly and flatten below 0.1 M$_\odot$ in order to explain the paucity of
observed brown dwarfs compared to the predictions (Basri \& Marcy 1997; Larson 
1998; Elmegreen 2000b; Kroupa 2002).  

Empirical studies of the IMF have been conducted recently in different
environments including clusters and associations in the Galaxy and the
Magellanic Clouds (von Hippel et al.~1996; Hunter et al.~1997; Hillenbrand 1997;
Massey \& Hunter 1998). Although all of these studies support a Salpeter IMF
with a slope in the neighborhood of $\sim 1.35$ for stars more massive than a
solar mass, some different slopes have been suggested (Scalo 1998; Massey \&
Hunter 1998). A top-heavy IMF for the first stars has also been suggested by a
number of authors (e.g., Larson 1998; Bromm et al. 2001a). Two functional forms
were proposed by Larson (1998).  
\begin{equation}
{dN \over d \log{m}} \propto (1 + m/m_l)^{-1.35}~~,
\label{larson1}
\end{equation}
and
\begin{equation}
{dN \over d \log{m}} \propto m^{-1.35} \exp{\{-m_l/m\}}~~.
\label{larson2}
\end{equation} 	
These both  approach the power law with a Salpeter slope
at high masses and fall off at the low-mass end.

Eq. \ref{larson1} falls off asymptotically to a slope zero at the low end. Eq.
\ref{larson2} has a peak at $m = m_l / 1.35$, and falls off exponentially with
increasing negative power at lower masses. The characteristic mass scale, $m_l$,
has a value of 0.35 M$_\odot$ for the present-day solar neighborhood. This mass
function was used by Hernandez \& Ferrara (2001), who suggest that the IMF for
population III stars was strongly weighted towards high masses at redshifts $6 <
z < 9$. They infer a value for $m_l$ of 12.2 at redshift $z \sim 9$, and
metallicity [Fe/H] $^<_\sim -2.5$, based on number counts of metal-poor stars
from the HK survey (Beers, Preston, \& Shectman 1985; 1992). Both functions were
tested in this model and compared to the simple power law. The mass scale was
chosen to be a linear time dependent quantity of the form $m_l = 13.2 - 0.9 t$.
The range of the IMF is taken from 0.09 to 40 M$_\odot$.
 
Figure \ref{fig:1} shows the different mass functions explored in this study. The
first is the simple Salpeter function with index $=1.35$. The second is a more
moderate high-mass biased Salpeter-like function that shows a slower decline
towards high masses than the simple function, but still steeper than the
top-heavy functions of Larson.  The two functions from Larson behave differently
at low masses. While the first produces a fair amount of low-mass stars and
declines slowly at the massive end, while the second function produces very few
low-mass stars and peaks at an intermediate value before it declines
exponentially. 

\subsection{Initial Conditions}

The baryonic masses of globular clusters observed in the Galaxy range between
10$^4$ and 10$^6$ M$_\odot$, while the minimum total mass of dwarf galaxies in
the Local Group are $\sim 2 \times 10^7$ M$_\odot$ (Mateo 2000). Therefore, we
choose initial total cloud masses in the range (10$^5$-10$^8$) M$_\odot$. In
this mass range, the clouds are sensitive to stellar feedback, both in terms of
chemical enrichment or energetics. We start with gas in a single phase and
uniform density, and choose the volume of each cloud such that we maintain a
reasonable initial density in the clouds ($\sim 0.25$ particles cm$^{-3}$)
corresponding to a baryon surface density of about 0.01 M$_\odot$ pc$^{-2}$.
This condition is necessary in order to calculate the initial potential energy
of the cloud, and its lifetime prior to destruction by SN energy input.

The dark matter component of the clouds is taken to be $\sim 3$ times the baryon
mass, consistent with observations of rich galactic clusters (e.g., Mushotzky et
al. 1996). The baryonic component of the clouds are taken to be composed
initially of 77\% (by mass) Hydrogen, 23\% Helium, and a zero metallicity. The
first massive star forms at the core and initiates subsequent events of
SN-induced star formation. This method of star formation produces local
inhomogeneities in the clouds as soon as the first SN event takes place. Each SN
event will produce a group of stars reflecting the abundance pattern of that
particular SN progenitor, since the stellar yields of a SN are different for
different progenitor masses.

It has been shown (Audouze \& Silk 1995) that mixing timescales in the early
halo are sufficiently long that chemical inhomogeneities in the gas would not be
erased on the timescale over which early generations of stars form.  Therefore,
we do not allow mixing between shells. Still, we assume instantaneous mixing of
the shell material left behind after the explosion with the gas in the ISM.
 
Our clouds are self-enriched over a timescale of $\sim 1$ Gyr.  The timescale for
the formation of the early halo is estimated to be of the order of a few Gyr
(Bekki \& Chiba 2000), rather than 10$^8$ yrs as suggested by ELS, although
most of the processes that contributed to the formation of the earliest
generations of stars will take place in the first 0.5 Gyr after the initiation
of the first enrichment events.

\section{Results and Discussion} 

\subsection{Stability of the Clouds}

Ignoring any tidal effects, heating and cooling, and any inhomogeneities in the
gas, and taking the energy input per SN event $E_{SN}$ to be $10^{51}$ erg, we
find that under the adopted initial conditions, clouds with total mass 
$^<_\sim 3\times 10^6$ M$_\odot$ (baryon mass $^<_\sim 7\times 10^5$ M$_\odot$)
do not survive the first SN explosion, since their total
gravitational potential energy is less than $E_{SN}$.  For more massive clouds,
the survival time depends on the SFR chosen and on the form of the IMF.

The initial density has a large effect on the stability of a cloud, and we must
keep in mind that star- forming regions become cooler and denser as time
passes.  As they form stars, the clouds are heated by the thermal input from
massive stars.  Therefore, such a simple treatment can only give a rough
estimate of the fate of these clouds.  Table \ref{table:1} shows the survival
times of our model clouds against destruction, as a function of initial mass,
for three different mass functions.

\begin{deluxetable}{lrrr}
\tablenum{1}
\tablewidth{0pt}
\tablecaption{Survival Times of the Clouds}
\tablehead{
\colhead{M$_{cloud}$ (M$_\odot$)}  & $m^{-1.35}$ & $(1 - e^{-m/2})m^{-1.35}$ & 
$m^{-1.35}e^{-m_l/4}$ \nl
}
\startdata

$3 \times  10^6$  & $> 10$ Gyr& - & - \nl
$5 \times  10^6$ & $> 10$ Gyr& 9 Myr & 8 Myr \nl
$1 \times 10^7$ & $> 10$ Gyr& 80 Myr & 12 Myr \nl
$1 \times  10^8$ & $> 10$ Gry& 200 Myr & 90 Myr \nl
 
\enddata
\label{table:1}
\end{deluxetable}

It is clear from the table that the fate of clouds which survive the first
explosion varies greatly depending on the type of the IMF chosen. The Salpeter
function produces very few massive stars, which allows enrichment to continue in
all clouds more massive than $5\times 10^6$ M$_\odot$ for times longer than 10
Gyr. At the same time, clouds as massive as 10$^8$ M$_\odot$ do not survive more
than $\sim 0.1$ Gyr with a top-heavy IMF. This leads to the suggestion that even
if the first stars were formed according to a top-heavy IMF, subsequent star
formation must have involved an IMF similar to the present day mass function.  

This result is in agreement with recent studies concerning the type of IMF
expected to govern mass formation in the past and present (Larson 1998;
Hernandez \& Ferrara 2001; Nakamura \& Umemura 2001; Nakamura \& Umemura 2002).
Larson (1998) suggested the existence of a mass scale in the star-formation
process which varies with cloud conditions such as pressure and temperature, and
therefore is time dependent. If the temperature in star-forming clouds was
higher at early times, this will have the effect of increasing the mass scale
and the relative number of low-mass stars formed at early times will decrease.
He also suggested another possible form of IMF with a universal power-law form
at large masses, but that departs from this power law below a characteristic
time-dependent mass scale.

Empirical evidence has supported this picture, with a power-law form of the IMF
down to one Solar mass (see \S 2.3). Recent chemical evolution models have
tested the time varying IMF in our galaxy as a solution to the G-dwarf problem
(e.g., ~Martinelli \& Matteucci 2000). These authors have found that an IMF with two
slopes, and a time- dependent shape at the low mass end, is required to
reproduce constraints other than the G-dwarf problem. They also show that such a
function is still not able to reproduce the properties of the Galactic disk and
suggested the inclusion of radial flows to reproduce the observed trends.
Invoking different slopes of the IMF in order to explain the G-dwarf problem is
a long-standing way to address this problem which has fallen out of favor (see
above). More recently, however, dynamical effects, both of the infall of
low-metallicity gas, and radial outflow of hot metal-rich gas (e.g., Samland et
al. 1997) have provided a somewhat more plausible mechanism to produce these
effects. Nevertheless, the question still remains as to whether the mass
function has varied with time and needs to be addressed further by including
dynamical effects in the models of galactic evolution. For the purposes of the
present study a simple time-dependent IMF is adequate. 
	
Our model is consistent with the clouds being the environment out of which the
globular clusters first formed.  It is  necessary for 
the efficiency $\epsilon$ of star formation be very high for
globular cluster formation.  The gravitational binding of the star
cluster that is formed must therefore dominate, because gas expulsion 
did not lead to the cluster's disruption.
The cloud mass which survives disruption ($\sim
7 \times 10^5$ M$_\odot$) is consistent with observed globular cluster masses,
particularly if some evaporation has occurred to the present time. In our model,
sufficiently massive GCs could have been self enriched. Indeed, a recent more
detailed model by Parmentier (2004) for the formation of globular-cluster
systems suggests that they were indeed self enriched.  A caveat, however is
that if  the efficiency $\eta $of star formation is very high, as e.g. for
globular cluster formation, the gravitational binding of the star
cluster that is formed must dominate because gas expulsion 
did not led to the cluster's disruption.

Alternatively, globular clusters could have been formed initially in more
massive stable systems such as dwarf galaxies. The homogeneous chemical
composition of stars within most globular clusters seems to suggest that they
formed out of gas that was already pre-enriched and well mixed inside structures
that were more stable against destruction. The globular cluster population of
the halo might then be explained as the result of accretion events during tidal
interactions with other galaxies. This was suggested previously by Armandroff
(1993) and Larson (1999).  

As mentioned previously, our calculation of the lowest mass of a cloud that can
survive the first SN event is very simplistic and should, perhaps, not be taken
too seriously on this issue of globular cluster formation and survival. Among
other things, our model ignores hydrodynamical effects and
inhomogeneities/fragmentation in the gas. More precise understanding of these
first clouds is required before definite conclusions can be made about the
formation of globular cluster systems. 

In clouds that do survive SN explosions, star formation halts
when there is not enough gas to form the shells that trigger the process.
Hence, the SN-induced star formation method used here is capable of 
self regulation.  If these clouds are allowed to accrete gas
from any other source, another sequence of star formation is possible.  This is
consistent with the observations that suggest that the Sgr dwarf galaxy has had
an episodic star formation history (Mighell et al. 1999).

\subsection{Stellar Abundances and Metallicities}

The self-enriching fragments of gas, in which the first stars are formed, will
eventually be destroyed either by internal stellar feedback or by external
effects such as tidal forces. Star formation is then halted. The extremely
metal-poor stars left behind contribute to the halo field population, and
contain information about the nucleosynthesis of elements in zero-metallicity
stars, their IMF, and the environment in which the first clouds formed. Thus, we
next compare the results of our calculation to the observed abundance trends in
the Galaxy.

\subsection{Metallicities and Ages of Stars}  

The metallicities of stars are calculated in this model from the metallicities
of the shells formed by SN events. (In this study, metallicity always refers to
the value of [Fe/H].) The first star that explodes is a metal-free Pop-III star.
Its ejecta depends on its initial mass. The ejecta will mix with a given amount
of primordial gas to determine the metal composition of second-generation stars.
Figure \ref{fig:2} shows the metallicities produced in shells of different
progenitors as a function of time. They are shown for four different progenitor
masses: 13, 20, 30 and 40 M$_\odot$. The metallicities produced at early times
range from [Fe/H]$ \sim -2.5$ to values well below [Fe/H]$\sim -4$ for the most
massive progenitors of $\sim 40$ M$_\odot$. This implies that it might be
possible to form Pop-II stars with [Fe/H]$ < -4$. This is in agreement with the
recent discovery of a low-mass star with an iron abundance of [Fe/H] $= -5.3$
(Christlieb et al. 2002). 

Figure \ref{fig:2} also shows that SNeII      produce shells that do not exceed
an initial metallicity of [Fe/H] $^<_\sim -2.5$. Note, that this uper limit of
[Fe/H] $^<_\sim -2.5$ is based upon a complete mixing of the supernova ejecta
into the shell. If this mixing were less efficient, for example by the loss of
ejecta from the cloud in a jet, this limiting metallicity would be appropriately
reduced. The fact that this limit is the value of metallicity at which McWilliam
et al.~(1995) observed a shift in the slope for relative abundances
vs.~metallicity suggests that mixing of ejecta with the shell must be rather
efficient. This value of metallicity separates two different stages of chemical
evolution, the era of inhomogeneous composition dominated by SNeII,
followed by a stage of rapid iron enrichment by type-Ia SN. As long as the
mixing of the ejecta is efficient, this also supports the idea that only a few
SN events are required to raise the heavy-element abundances from zero to
the values observed in the most metal-poor stars in the halo.  

In Figure 3 we show the age-metallicity relation for stars in a cloud of
initially primordial gas, and assuming a Salpeter mass function. This cloud was
allowed to evolve up to 10 Gyr. The ejecta of type-Ia SN contributed after a
delay time of 1 Gyr. Each point on the figure represents a shell that formed
stars. The mean value of [Fe/H] increases with time, as expected. Stars of the
same age show a dispersion in metallicities that increases for older stars. This
is not predicted by a simple one-zone model that assumes complete mixing, and is
a result of the star formation mechanism applied in this model, i.e. that stars form
out of the material ejected in Type-II SN events. This will allow for a variety
of chemical composition depending on the progenitor masses and metallicities. 

This large scatter, decreasing with time, can be explained by the fact that
stars are formed only during SN events. Therefore, stars are the result of the
mixing of SN ejecta with a limited amount of gas. This produces stars of
different elemental ratios even if they form at the same time, since the SN
ejecta are dependent on both the initial mass and metallicity of the progenitor. 

The most-massive SNe would contribute first and reflect their yields in the
first Pop-II stars to appear, then at later times the products of the
less-massive SN are expected to appear with different elemental ratios (WW95).
This does not necessarily mean that the first and most massive SNe will produce
the lowest metallicity Pop-II stars. It is the lowest mass SNeII that
produce shells as low in metallicity as [Fe/H] $\sim -4$, while 20 to 30
M$_\odot$ progenitors produce higher metallicity shells. 

When enough time has passed for the ISM to mix on a large scale, the elemental
ratios are expected to reflect an average of the ejecta of SN events of
differing masses, including those of type-Ia. The steep rise at early times
reflects the dominance of SN individual events with different effective yields,
and a high efficiency for the enrichment of the metal-poor ISM. The gradual
increase at later times is due to the slower enrichment process involving a
well-mixed ISM. It is clear from the scatter in the figure that a simple
age-metallicity relation probably does not exist at early times.

\subsection{ The Metallicity Distribution Function (MDF)}

Carney et al. (1996) compared the kinematics and chemical properties of
metal-poor stars as a function of
distance from the Galactic plane. They found that stars 
closer to the plane exhibit a slightly more metal-rich mean abundance, 
$\langle$[Fe/H]$\rangle \approx  -1.7$, while those farther from the plane have  
$\langle $[Fe/H]$\rangle \approx -2.0$. This was  postulated to
result from the existence of two 
discrete components in the halo. The flattened, inner component was probably 
formed in a manner resembling the ELS model, while the more spherical, outer 
halo had a large contribution from nearby smaller systems similar to the 
Sagittarius dwarf galaxy or other proto-dwarf galaxies. 
This is in agreement with the hierarchical CDM model (Kauffman, White, \& Guiderdoni 1993).
The halo MDF is therefore generally believed to be the result of
combining the distribution functions of individual proto-dwarf galaxies
accreted by the Galaxy (Cote et al.~2000; Scannapieco  \& Broadhurst 2001, and
references therein). 

Although many recent studies of the formation of the halo are supportive of
such a scenario (Cote et al.~2002; Dinescu 2002; Brook et al.~2003), when
comparing the observed features of present-day dwarf galaxies to those of the
(presumably) accreted stars and clusters, one finds inconsistencies. The so-
called {\it satellite catastrophe} (Mateo 1998; Moore et al. 1999), the {\it
angular momentum catastrophe} (Navarro \& Steinmetz 2000), and comparisons of
abundances and ages in large and dwarf galaxies, lead to the conclusion that
present day dwarfs are probably not the major building blocks of large galaxies
like the Milky Way (Shetrone, Cote \& Sargent 2001; Baldacci et al. 2002;
Kauffmann et al.~2003; Tolstoy et al. 2003). This is not necessarily in
disagreement with hierarchical galaxy formation theories. One can still argue
that the satellite systems that built large galaxies may have been different
from present-day dwarfs, and that these systems are not visible today (Steinmetz
2003). For a more detailed discussion, see Tosi (2003). 
   
The observed MDF of metal-poor stars in the Galaxy shows a relatively broad
peak with a maximum at [Fe/H] $\sim -1.6$, and a smooth tail extending to values
of   [Fe/H]$^<_\sim -3$.  In contrast, the metallicity distribution of disk stars
shows a more localized peak with a sharp cutoff (Ryan \& Norris 1991a;
1991b).  This suggests that the environment in which the disk formed was
probably more uniform than that which formed the halo stars. 

A number of authors (e.g., Norris 1999; Chiba \& Beers 2000) have suggested that
there might be a "bump" in the halo MDF at lower metallicities. A feature not
expected from the simple model, this bump might be suggestive of non-uniform
enrichment at early times. In this picture, pre-Galactic clouds are assumed to
be the building blocks of the halo. Therefore, their metallicity distributions
will contribute to the total MDF of the halo. The MDF is expected to vary from
cloud to cloud depending on the initial conditions and the duration of chemical
enrichment.

As a crude approximation to the effect of an expected range of combined MDF's
we made a simple study of the MDF's of four different clouds, each evolved
with a different IMF.  We did this under 
the assumption that all these clouds eventually 
disperse and contribute to the field of the halo.
The time evolution of the MDF was followed up to the time of
each cloud's destruction. In the case of the Salpeter mass function, 
for example, the
evolution was followed up to 5 Gyr and was found to saturate at 1 Gyr.  

Figure \ref{fig:4} shows the results for the four clouds, where the type of mass
function chosen for each cloud is shown on the figure (all MDFs were normalized
to unity). The panels shown in Figure \ref{fig:4} indicate that the chemical
enrichment in the first cloud, with a simple power-law mass function, is much
slower than in the other three clouds with mass functions biased toward massive
stars. Therefore the timescales chosen for this figure are different, since the
enrichment is slower and with the short timescales used in the other three
panels, one would not be able to see the full evolution picture in this cloud. 

At the beginning all of the clouds show a dispersion, with the
majority of stars at metallicities below [Fe/H] $= -5$.  This describes the
first stars that form from the shells of the most massive progenitors with $m >
35$ M$_\odot$.  This could also be concluded from Figure \ref{fig:2}, which
shows that only these massive stars form such low-metallicity shells. 

As time goes on, less-massive stars will contribute shells in the metallicity
range $-3 \le $[Fe/H]$ \le -2.5$. These produce a rapid increase in the numbers
of stars in this metallicity range. Moreover, these stars should greatly
outnumber the first stars formed from higher-mass explosions. As soon as stars
with $m \le$20 M$_\odot$ start to contribute, shells with metallicities down to
[Fe/H] $\sim -4.5$ form and a dispersion is seen once again. This dispersion
eventually fades out as the inhomogeneities start to disappear and the stars
become more metal rich. The final stage shows a peak at a different value of
metallicity for each cloud. This illustrates the different enrichment produced
by the different mass functions.

The first cloud, with a simple mass function, achieves a peak at [Fe/H] $=
-2.7$, a value consistent with the calculations of Ryan et al. (1996) and
Nakasato \& Shigeyama (2000) for the expected metallicity of Pop-II stars formed
in the shell of a typical SN expelled into primordial gas. The other clouds
produce higher metallicity peaks, corresponding to the larger number of massive
stars that they produce. These peaks have the values [Fe/H]$= -2.6, -2.5$, and
$-2.2$, respectively. The latter corresponds to the fourth mass function shown
in Figure \ref{fig:1}, which produces the largest number of massive stars. 

In Figure \ref{fig:5} we illustrate the sum of the final stages of all four clouds. This
produces a broader distribution extending from about [Fe/H]$= -3.2$ to about
[Fe/H] $= -2.2$ and peaked at a value of [Fe/H] $= -2.6$. The MDF observed in
the halo shows a yet broader distribution, than that indicated on Figure
\ref{fig:5}. This is probably due to the large number of different systems that
contributed to it, and a higher value for the maximum mass indicating the
contributions from systems that may have produced stars from pre-enriched gas.

\section{Relative Elemental Abundances} 

\subsection{ The Observed Trends in Halo Stars}

Very metal-poor halo stars show a great diversity in their absolute elemental
abundances and can show considerable scatter (up to a factor of 100) in the observed
abundances of heavy elements relative to Fe. (Ryan \& Norris 1991; Gratton \&
Sneden 1994; McWilliam et al. 1995; Ryan et al. 1996; McWilliam 1998; Sneden et
al. 1998; Norris, Ryan, \& Beers 2001, Carretta et al. 2002). At higher
metallicities the scatter gradually decreases until this ratio terminates at a
value that corresponds to an average over the IMF of the element to iron ratios
of the stellar yields.  

McWilliam et al.~(1995) detected a shift in the slope of the abundances of the
elements Al, Mn, Co, Cr, Sr, and Ba relative to Fe below [Fe/H]$\sim -2.5$. They
suggested that this unusual chemical composition must have originated from the
fact that supernova yields must have changed with time, or in other words, are
metallicity dependent. These results were confirmed later by Ryan et al.~(1996)
who also showed that scatter in the abundance ratios increases with decreasing
[Fe/H]. 

Under the assumption of efficient mixing of the supernova ejecta with the
clouds, these observations support the idea that the most metal-poor stars
exhibit the ejecta of very small numbers of supernovae. The
metallicity-dependent yields used in our present model, together with the unique
method of SN-induced star formation, are capable of explaining these trends. The
observed relative abundances in the halo are fully reproduced with this model
for several elements. These results are summarized and discussed in the
following sections.

\subsection{Overview of Model Predictions}

To obtain a picture of the early elemental abundance evolution, simulations were
run for 5 Gyr using a massive cloud ($10^7$ M$_\odot$), and a Salpeter initial
mass function. After 1 Gyr, SN of type-Ia were allowed to contribute their
ejecta. The elemental abundances were calculated and recorded for every shell
that forms during the simulation. The model produces a small number of stars at
very low metallicities, [Fe/H]$ < -3$, showing a considerable spread in [X/Fe]
ratios, ranging from less than 0.5 dex in the case of Ca and Ti to as much as 
1 dex for Mg (Cohen et al. 2004) [see however Arnone et al. (2004)
who obtain almost no dispersion for Mg]. 
The observed scatter of the abundances for model stars is given by the
spread in metallicities of the SN models. Therefore, the large scatter in the
abundance ratios observed in low-metallicity stars is 
inherently reproduced by the stochastic nature of this model.

Local inhomogeneities start to disappear at  $-3.0 < $ [Fe/H] $ < -2.5$,
corresponding to $\sim  0.1$ Gyr from the onset of star formation. At this
stage most of the massive SNe have contributed, and their ejecta have already
mixed with the gas in the ISM. This metallicity range marks the end of the
early phase and the beginning of the transition to the well-mixed phase. At
this stage, the gas becomes more metal-rich.  Hence, newly formed stars will no
longer exhibit abundance patterns of a single SN, but rather an average over
the IMF of the supernovae that contributed to the enrichment of the local ISM.
These values resemble the predictions of a simple one-zone model 
in which stars are assumed to form out of the well mixed gas in the ISM. The
spread in the relative abundances decreases gradually, reflecting the ongoing
mixing process as more SNe pollute the ISM, and the values of [X/Fe] approach
the solar value.

\subsection{[$\alpha$/Fe]}

Figure \ref{fig:6} shows  [$\alpha$/Fe]
vs. [Fe/H] for several alpha elements calculated by 
a straightforward application of the model (small dots). These are compared with two
sets of observational data, Ryan et al.~(1996) (triangles) and Norris et
al.~(2001) (circles). The observed values show an over-abundance relative to
solar, and a scatter that is significant at all values of [Fe/H], larger at the
lowest abundances. There is also a peculiar trend in the data to show some stars
which appear to have increasing [$\alpha$/Fe] with decreasing [Fe/H] which shows
up as an extension in the data. This behaviors is apparent in the abundances of
Mg, Ca, and Si. Ti, on the other hand, shows a scatter that is almost constant
over the metallicity range. 

The over-abundance in alpha elements relative to iron compared to the Sun can
be understood as the result of the dominance of SNeII         at the time, since
they are believed to be the major production sites for alpha elements. This also
indicates that the corresponding stars formed within the first Gyr of the onset
of significant star formation. After 1 Gyr, type-Ia SN events are expected to
shift the trends with large amounts of Fe. Therefore, the ratios decline
gradually towards solar metallicities. 
The so-called plateau for [$\alpha$/Fe]
ratios at low metallicities 
has been shown (Arnone et al. 2003; Cayrel et al. 2004; Cohen et al. 2004) 
to have less dispersion  (at least for some elements) than previously
thought (Ryan et al.~1996; McWilliam 1997; Norris et al.~2001). 
This dispersion can set strong limits on the nature of the
first SN, and the nature of the enrichment process when compared to the
predictions of chemical evolution models.

More recent studies of the abundances of metal-poor stars show similar trends in
general (Carretta et al.~2002; Fulbright 2002; Ivans et al.~2002;). In addition,
all three studies report the detection of stars with low alpha-element
abundances relative to iron, in contrast to the majority of metal-poor stars. To
explain these observations, they suggest that these stars probably witnessed the
enrichment of type Ia SN events (Ivans et al.~2002), owe their origins to lower
mass stellar systems with slower enrichment histories (Carretta et al.~2002), or
were formed in the Galaxy but at times of nonuniform mixing where a mass
function that produces less high mass stars (which are responsible for producing
Mg as less massive SNII produce the heavier alpha elements) would be responsible
for the low-alpha stars. This is an interesting phenomena that needs to be
included in a complete picture for the chemical evolution of the Galaxy, once
enough data is available, since it does have the potential to set limits on the
type and number of SN events that took place in the environment from which these
stars formed. 

Carretta et al.~(2002), Arnone et al.~ (2004), Cayrel et al.~2004. and Cohen 
et al.~(2004)
have   analyzed a sample of extremely metal-poor stars, down to 
low metallicities, [Fe/H]$\ge -3.0$. They find that the scatter in the abundance 
ratios [X/Fe], is surprisingly small, even at the lowest metallicities, compared 
to what is expected from a stochastic early evolution. They suggest several 
ideas to reduce the scatter predicted by chemical evolution models, including 
effective mixing to maintain the chemical homogeneity in the clouds, limiting 
the mass range of the IMF, introducing a first generation of very massive stars 
($> 100$ M$_\odot$), and relaxing the closed-box approximation.  

As seen in Figure 6, our most naive model 
reproduces the basic observational trends for alpha elements relative to iron.
However, the model tends to 
overestimate the scatter for alpha elements, and exhibits over all lower
values than expected for [X/Fe] ratios.  This suggests an  over production of Fe.
This is most evident in Mg and Ti, which show  mean values
of [X/Fe]  that are 0.5 dex lower than
the data all the way down to the lowest metallicities. Ca and Si show better 
agreement below [Fe/H] $< -3$, but also decline and reach subsolar values as
[Fe/H] increases. The suggestion by Timmes et al.~(1995) of reducing the amount
of Fe produced by massive stars seems to be required, especially in the cases of
elements Mg and Ti. This indicates the uncertainty in the Fe yields, which is
representative of Fe-group elements in general, and will be discussed in the
next section. The overproduction of Fe can also be attributed to the
contributions of SNIa. A delay time longer than 1 Gyr may be required. Although
this is a possibility, it still seems more likely that the Fe yields from SNeII        
are responsible, since the apparent underproduction of alpha elements seen
in our calculation is not seen for Fe group elements as well.
  
Figure 7 shows the results of our model when the Fe yield is reduced by a factor
of 2. The agreement between the data and the calculation in this case is much
improved. We even produce the bifurcation of trends, e.g. in [Mg/Fe]
and [Si/Fe] there are two trends in the data.  One is a flattening at
[X/Fe]$\sim 0.5$, and another a linear increase in [X/Fe] with decreasing metallicity.
 In our model, these two trends are an artifact of
which type of star produced the first supernova. The metallicity-dependent
yields of Ca reproduce the values of [Ca/Fe] and the scatter at low
metallicities very well, and show a flattening with decreasing metallicity.
This was also observed in the samples analyzed by Carretta et al.~(2002), 
Cayrel et al.~(2004) and Cohen et al.~(2004). 
Carretta et al. (2002) also suggest that the scatter in [Si/Fe] ratios at low
metallicity is due to uncertainties in the observational techniques used for this
element.
Our results for the alpha element Ti shows lower values than the data and
rather little dispersion. This is consistent with the most recent observations 
(e.g. Cayrel et al. 2004; Cohen et al. 2004).  However, as for any element, 
the less abundant it is, the more uncertain is its production rate,
 as calculated from supernova models.  This is surely the case tith Ti.

Previous studies (e.g., Argast et al. 2002) have found that such stochastic
models tend to predict too much scatter (e.g. for O and Mg) and fail to
reproduce the trends of other elements. Our model is also based upon a
stochastic picture and
also can overestimate the scatter in Mg compared
to the most recent observations unless a modification in the yields is adopted as in
 Argast et al.~(2002).
This will be discussed in more detail in \S 4.5. 

\subsection{[Fe-peak / Fe]}

For Fe-peak elements, the element abundance ratios relative to iron are close to
solar down to a metallicity [Fe/H]$ ^>_\sim -2.5$. That is, except for Mn,
which is underabundant in all halo stars. Below this ([Fe/H]$^<_\sim -2.5$), the
behavior changes (McWilliam et al. 1995; Ryan et al. 1996; McWilliam 1997;
Norris et al. 2001). Both Cr/Fe and Mn/Fe decrease to sub-solar values, Co/Fe
increases to super-solar values, and Ni maintains an average value slightly
above solar. The scatter increases towards lower metallicities.
The correlation seen between Cr/Fe and Mn/Fe ratios (both
increasing with increasing metallicity), was speculated by Carretta et
al.~(2002) to be an indication of the absence of very massive stars ($m \ge 100$
M$_\odot$), since they argue that these 
stars would under-produce nuclei with odd nuclear charge.
Therefore, [Mn/Fe] would decrease with metallicity while [Cr/Fe] would remain
approximately constant. They suggested an explanation for the trends of
these ratios decreasing with decreasing [Fe/H] as due to variation in mass cuts
in SNeII      events as a function of progenitor mass. The observed trends can
be reproduced if the mass cut is smaller for larger progenitor mass, which
presumably were more common at lower metallicities.

Figure 8 shows the calculated values of relative abundances for Cr, Co, Mn
and Ni.   The observed ratios of Mn and of Ni are reproduced well by the
metallicity- dependent yields of WW95 in this model. Cr, on the other hand,
does not seem to produce satisfactory results. The stars produced at very low
metallicity show a small over-abundance relative to solar, contrary to the
observed under-abundance.   The excess in Co/Fe was produced fairly well with
the metallicity-dependent yields below [Fe/H] $= -2.5$ if the contributions
from SN  more massive than 40 M$_\odot$ are excluded (see section 4.5). This
behavior of Co at the lowest metallicities has not been produced by any
previous models, and no known stellar yields were able to predict it even by
modifications in explosion parameters of SN-II models (e.g.~Nakamura et al.
1999).

Our calculations raise the questions as to how well the nucleosynthesis of
massive stars is understood, especially for progenitors of low metallicity. The
dependence of the yields of these elements on the position of the mass cut and
degree of mixing for material ejected from SNe is one of the primary sources of
large errors in calculated abundance yields. Mn and Cr are produced mainly
during explosive Si burning, and therefore have a complicated dependence on the
progenitor mass. Efforts to understand the abundance trends of iron-peak
elements, which are formed close to the mass cut, have focused on the possible
alpha-rich freeze-out (McWilliam et al. 1995), the location of the mass cut, and
the dependence of yields on the star mass, metallicity, and neutron excess
(Nakamura et al. 1999). Our model predicts a scatter in the ratios for Fe-group
elements that is larger at very low metallicities than the observed values.
Mostly this large scatter is contributed from the products of SNe more massive
than 35 M$_\odot$. This may suggest that these stars do not contribute and end
their lives as a black hole (see \S 4.5). This could also be attributed to
the several approximations made in the model, especially the exclusion of any
overlap or mixing among SN shells.

\subsection{Contributions from Different Supernovae} 

The ejecta of the most massive stars are expected to dominate the earliest
phases of the chemical evolution of each cloud. These stars produce values of
[Fe/H]$< -2.5$ in their shells. According to WW95, at zero metallicity, the most
massive stars (35 M$_\odot< M < 40 $M$_\odot$) produce metallicities lower than
that observed while the lowest observed metallicities of [Fe/H] $\sim -4$ are
produced by progenitors of masses less than 20 M$_\odot$. These various
behaviors of SN with different progenitor masses is responsible for the
dispersion in metal abundances at low metallicities. Studying the dispersion in
the relative abundances of elements in the halo will allow for constraints to be
set on chemical evolution models. It can help us understand the nature of the
first SNe, and the inhomogeneous enrichment processes that took place. It also
constrains the nature of the IMF and star formation rates at the corresponding
times.

Figures \ref{fig:9} to \ref{fig:12} show the contributions to the elemental
abundances by different progenitor masses down to metallicities of [Fe/H]= $-5$
and up to $-2.5$. This range represents the early stages of chemical evolution in
the Galaxy for which our model is applicable. Once [Fe/H] $> -2.5$, the
contributions from SNIa pollute the ISM and the contributions from SN of Type II
become less dominant. For these figures, we have run the simulation allowing
only a range of progenitor masses to contribute its ejecta. The mass ranges are
shown on the figures. The same is shown for Fe-group elements in Figures 
\ref{fig:13} to \ref{fig:16}.  

The main feature that can be seen in all eight figures is that the progenitors
responsible for producing extremely low metallicities in the Galaxy are either
more massive than 35 M$_\odot$ or less massive than 20 M$_\odot$. Figure
\ref{fig:2} shows that these extremely low-metal stars are only produced at the
earliest times, and therefore, are probably the products of Pop-III stars. Also,
it is the the high-mass progenitors $m^>_\sim 40$ M$_\odot$ that produce shells
with metallicities way below [Fe/H] = $-5$, while the progenitors with $m ^<_\sim 20$
M$_\odot$ can only produce shells with metallicities down to [Fe/H]$^>_\sim -4$.
Therefore, the fact that the ejecta of the high mass progenitors is missing at
low metallicities, while the current belief is that they have been available in
the halo from the beginning, can definitely be attributed to a lack of data. And
while the search for lower metallicity stars still continues, this will remain
an open question. 

If the extremely low-metal stars are never found in our Galaxy, one can suggest
that the first stars that formed in pre-Galactic clouds formed out of
pre-enriched gas, and that there are no Pop-III stars in our Galaxy. Or that
stars more massive than 35 M$_\odot$ do not contribute their ejecta, and
instead, end their lives as black holes. 
   
The missing ejecta of SN events with progenitors less massive than 20 M$_\odot$
 in the data, at low metallicities, may be suggestive of the fact that such stars are
not produced at the earliest times, in support of the theory that a top-heavy
IMF dominated star formation in pre-Galactic clouds. Another feature in the
figures supporting this idea is that for several elements (Mg, Cr, Co, Ni and
Mn), the calculated relative abundances that lie outside the observed ranges at
metallicities [Fe/H] $> -4$ are either contributed partially (as in the cases of
Ni, Cr, and Co),or totally (as in the cases of Mg and Mn) by stars less massive
than 25 M$_\odot$. This is suggestive of the fact that SNe that 
contributed before [Fe/H]
$\sim -2.5$  were all massive. In the cases of the elements Ni, Co and Cr, we can also
see contributions outside the observed ranges of relative abundances coming from
stars more massive than 35  M$_\odot$. This also may imply that these stars do not
contribute as SNeII.  When these SNe at the low-mass and high-mass
end are not allowed to contribute at the earliest stages, the model
fits very well the recent observational data (Caretta et al. 2002; Cayrel et al. 2004;
Arnonne et al. 2004) which report less scatter than previous studies.

Again we need to keep in mind here that the
nucleosynthesis calculations of these elements are still not certain enough to
allow us to build accurate conclusions regarding their production. Therefore, we
do see here that there is a great need to study these early stages of chemical
evolution in more detail in order to be able to set limits on the details of the
production of these elements.

The calculated values from the high-mass progenitors do not fit the observed
values for Co even at higher metallicities, while the rest of the mass range
down to 10 M$_\odot$ reproduces the observations very well. This result for the
low-mass end is in contrast to what is found for the other elements Mg, Ti, Cr,
and Mn. In these cases, the low-mass progenitors produce values of the relative
abundances out of the range of the data. This may be suggestive of the fact that
the first stars were massive and that the SN that contributed before [Fe/H]$\sim
-2.5$ were all massive SN. The alpha-element enhancement observed for stars with
[Fe/H]$ <-2.5$ is indicative of a high-mass IMF in the first Gyr of the Galaxy's
formation. After this time, as type-Ia SN contribute their Fe ejecta, at [Fe/H]$
\sim -2.5$, the alpha-element ratios go down to solar values.

\section{Summary and Conclusion}

We developed a CDM-motivated model to simulate the stochastic early chemical
evolution of proto-Galactic clouds in the halo. This model is based upon a
SN-induced star-formation mechanism which keeps track of the chemical enrichment
and energy input to the clouds by Type II and Type Ia supernovae. An important
feature of this model is the implementation of metallicity-dependent yields for
all elements at all times, and the inclusion of finite stellar
lifetimes. These were found to have important effects on the metallicity
distribution functions for stars in the clouds, their age-metallicity relation,
and relative elemental abundances for of alpha- and Fe-group elements.  

The stability of these clouds against destruction was considered, and results
were found to depend upon the choice of initial mass function for
low-metallicity stars. Most clouds with initial masses similar to presently
observed globular clusters were found to survive disruption from the onset of
star formation, suggesting that these systems could have been responsible for
the formation of the first stars, and could provide an environment for the
self-enrichment of globular clusters. However, such clouds are only stable when
one assumes an initial mass function that is not too biased towards massive
stars, indicating that even if the first stars were formed according to a
top-heavy mass function, subsequent star formation may need to have proceeded
with a present- day mass function, or happened in an episodic manner.

An important conclusion of this study is that the dispersion of the metallicity
distribution function observed in the outer halo is naturally reproduced by
contributions from many clouds with different initial conditions. The scatter
in metallicity as a function of age for stars in our simulations is very large
at early times, implying that no age-metallicity relation exists in the very
earliest stages of galaxy formation.

Regarding observed elemental abundances, the observed scatter is reproduced
fairly well for most, but not all, elements. In particular, the predicted
relative abundances of alpha elements compared to iron can be made to agree with
the observed values down to a metallicity below [Fe/H] $\sim -4$, but only if
the total iron produced is reduced by a factor of two from our adopted WW95
yields. Similarly, iron group elements Cr, Co, Mn, and Ni all exhibit abundance
trends vs. [Fe/H] that are somewhat different than those observed. The simplest
explanation for both of these phenomena within our model is to move the mass-cut
for iron and iron-group elements outward in the stellar models for all stars
except those with masses in the range $25 < m < 35$ M$_\odot$. If yields are
restricted in this way a good fit to all elements can be obtained. We note,
however, that though this is a simple explanation, we can not rule out that
these trends could result from some of the previously noted shortcomings of the
model rather than a stellar evolution effect. For example, the reduced iron and
iron-group yields could perhaps equally well have been achieved by relaxing the
assumption of efficient mixing of the ejecta with the cloud.

A particularly reassuring result of the present study with regard to alpha
elements is that these models naturally predict the observed detailed trends.
That is, these models reproduce both stars which increase in [$\alpha$/Fe] with
decreasing [Fe/H] and stars which obtain a near constant value of [$\alpha$/Fe]
albeit with a significant scatter. In these models, this bifurcation of the
abundance distributions can be attributed to whether the initial supernovae
within the clouds were massive $m \sim 40$ M$_\odot$ progenitors or of lower
mass.

The contributions to the abundances from supernovae with different progenitor
masses and metallicity suggest that the low-mass end of Type-II SN was
probably absent at the very lowest metallicities, and that the upper mass limit
for the first stars that contributed to nucleosynthesis may be $^<_\sim 40$
M$_\odot$.

\acknowledgments
Work at Michigan State University supported by grants AST 95-29454, AST
00-98549, and AST 00-98508, as well as PHY 02-16783, Physics Frontier
Centers/JINA: Joint Institute for Nuclear Astrophysics, awarded by the National
Science Foundation. This paper represents the thesis work of L.
Saleh completed at Michigan State University in partial fulfillment of the
requirements for a Ph.D. degree. L. Saleh acknowledges the doctoral dissertation
completion fellowship granted to her by Michigan State University through the
research grants summarized above.

Work at the University of Notre Dame 
supported  by the US Department of Energy under Nuclear Theory grant
DE-FG02-95ER40934. 

-----------

\begin{figure}
\caption{
The different mass functions used in the model,  $\phi_1 = m^{-1.35}$, 
 $\phi_2 = (1-e^{-m / 2}) m^{-1.35}$ ,
$\phi_3 = (1 + m/ml)^{-1.35}$,    $\phi_4 = m^{-1.35} e^{- ml / m}$  .
}
\label{fig:1}
\end{figure}

\begin{figure}
\caption{ Metallicity enrichment vs. time in the first 2 Gyr for  shells containing  
progenitor stars  of various masses.
The progenitors shown have masses 13, 20, 30 and 40 M$_\odot$}
\label{fig:2}
\end{figure}

\begin{figure}
\caption{ Age-metallicity relation for stars produced by the model.
Note the large dispersion at early times.}
\label{fig:3}
\end{figure}

\begin{figure}
\caption[]{
 The metallicity distribution functions produced in clouds with 
four different initial mass functions as labeled. }
\label{fig:4}
\end{figure}

\begin{figure}
\caption[]{The sum of the MDFs for the four clouds shown in figure \ref{fig:4}}
\label{fig:5}
\end{figure}

\begin{figure}
\caption[]{The abundances of the alpha elements Mg, Ca, Si, and Ti. Data is from  
Ryan et al.~(1996)  (triangles) and Norris et al.~(2001) (circles).  The small 
(dots) represent the model stars}
\label{fig:6}
\end{figure}
%
\begin{figure}
\caption[]{The abundances of the alpha elements when Fe yields are reduced by a 
factor of 2. Data is from Ryan et al.~(1996)  (triangles) and Norris et al.~(2001) 
(circles).  The small (dots) represent the model stars}
\label{fig:7}
\end{figure}

\begin{figure}
\caption[]{The abundances of the Fe-group elements, Mn, Ni, Cr, and Co . Data  
is from Ryan et al.~(1996)  (triangles) and Norris et al.~(2001) (circles).  The 
small (dots) represent the model stars}
\label{fig:8}
\end{figure}

\begin{figure}
\caption[]{The contribution from different progenitor masses 
to the value of  [Ca/Fe]. The mass ranges are given in solar 
masses. The top left figure shows the whole range. Data is 
from Ryan et al.~(1996)  (triangles) and Norris et al.~(2001) 
(circles).  The small (dots) represent the model stars.}
\label{fig:9}
\end{figure}

\begin{figure}
\caption[]{The contribution from different progenitor masses 
to the value of  [Mg/Fe]. The mass ranges are given in solar 
masses. The top left figure shows the whole range. Data is 
from Ryan et al.~(1996)  (triangles) and Norris et al.~(2001) 
(circles).  The small (dots) represent the model stars}
\label{fig:10}
\end{figure}

\begin{figure}
\caption[]{The contribution from different progenitor masses 
to the value of  [Si/Fe].  The mass ranges are given in solar 
masses. The top left figure shows the whole range. Data is 
from Ryan et al.~(1996)  (triangles) and Norris et al.~(2001) 
(circles).  The small (dots) represent the model stars}
\label{fig:11}
\end{figure}

\begin{figure}
\caption[]{The contribution from different progenitor masses to the value of  
[Ti/Fe].  The mass ranges are given in solar masses. The top left figure shows 
the whole range. 
Data is from Ryan et al.~(1996)  (triangles) and Norris et al.~(2001) (circles).  
The small (dots) represent the model stars}
\label{fig:12}
\end{figure}

\begin{figure}
\caption[]{The contribution from different progenitor masses 
to the value of  [Cr/Fe].  The mass ranges are given in solar 
masses. The top left figure shows the whole range. Data is 
from Ryan et al.~(1996)  (triangles) and Norris et al.~(2001) 
(circles).  The small (dots) represent the model stars.}
\label{fig:13}
\end{figure}

\begin{figure}
\caption[]{The contribution from different progenitor 
masses to the value of  [Co/Fe].  The mass ranges are given 
in solar masses. The top left figure shows the whole range. 
Data is from Ryan et al.~(1996)  (triangles) and Norris et al. 
(2001) (circles).  The small (dots) represent the model stars}
\label{fig:14}
\end{figure}

\begin{figure}
\caption[]{ The contribution from different progenitor 
masses to the value of  [Mn/Fe].  The mass ranges are 
given in solar masses. The top left figure shows the whole 
range. Data is from Ryan et al.~(1996)  (triangles) and 
Norris et al.~(2001) (circles).  The small (dots) represent the 
model stars.}
\label{fig:15}
\end{figure}

\begin{figure}
\caption[]{ The contribution from different progenitor masses 
to the value of  [Ni/Fe].  The mass ranges are given in solar 
masses. The top left figure shows the whole range. Data is 
from Ryan et al.~(1996)  (triangles) and Norris et al.~(2001) 
(circles).  The small (dots) represent the model stars}
\label{fig:16}
\end{figure}

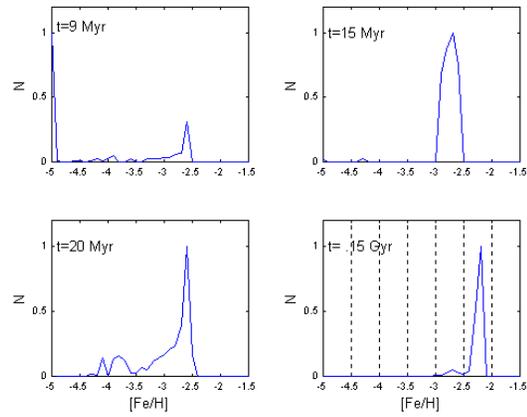
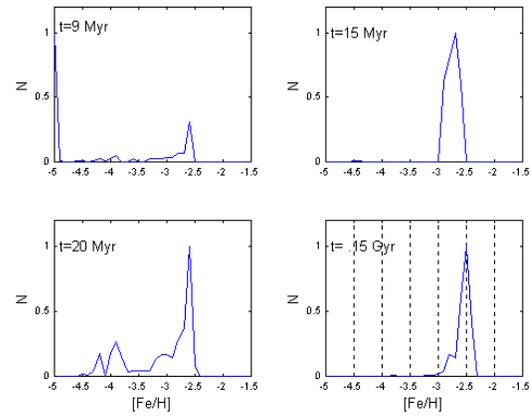
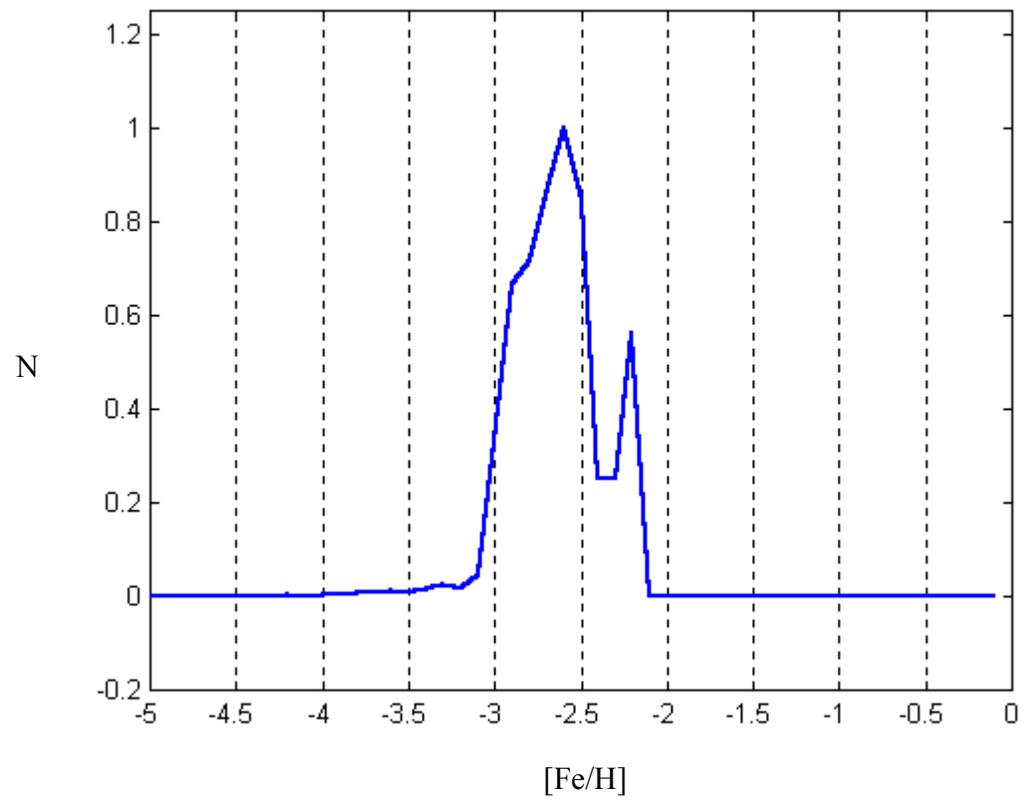
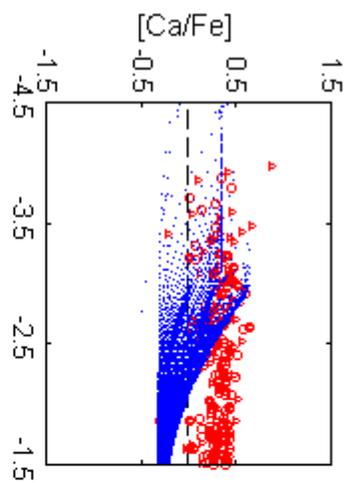
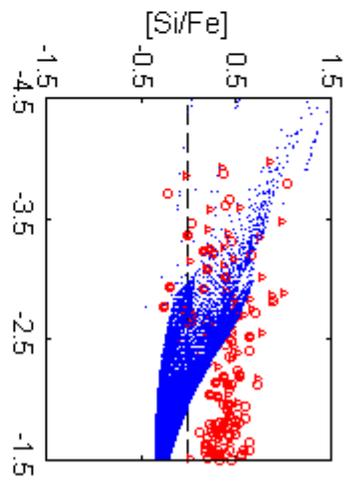
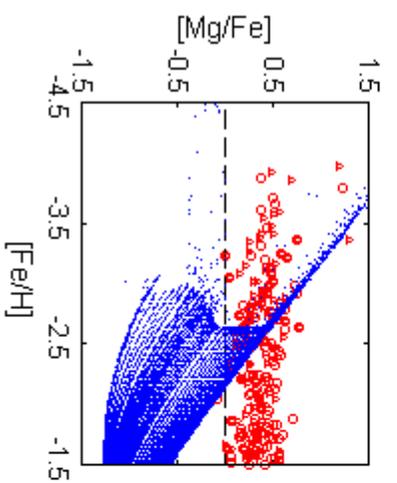
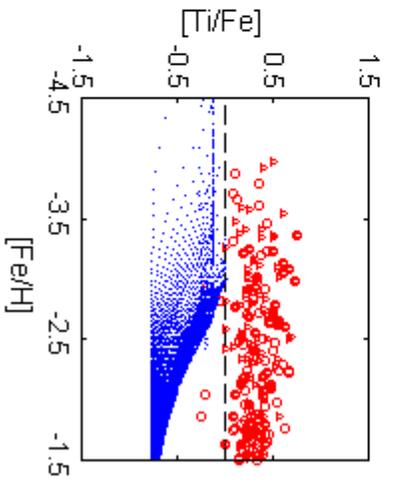
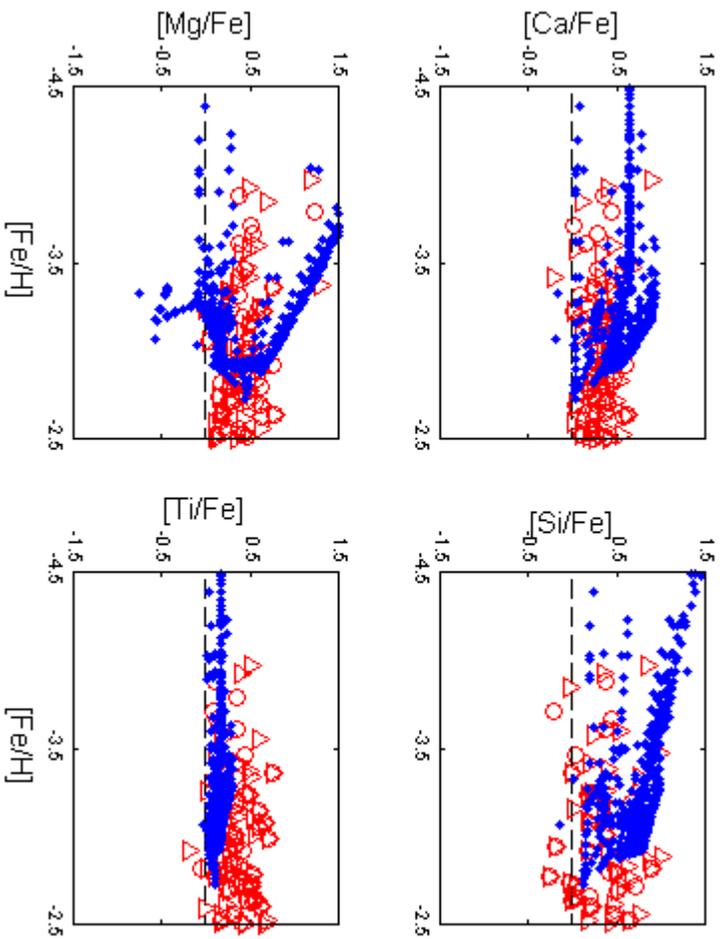
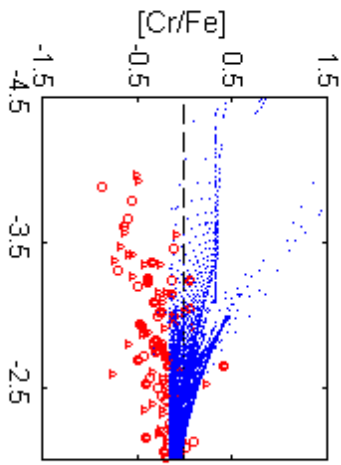
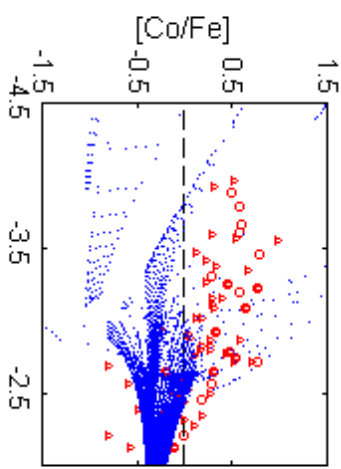
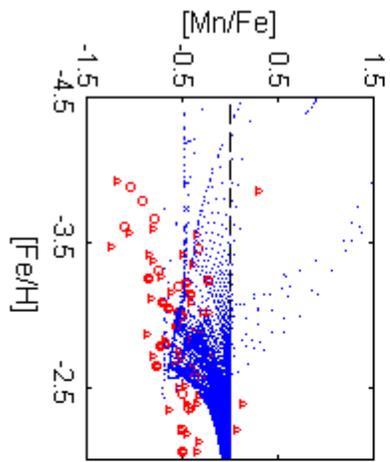
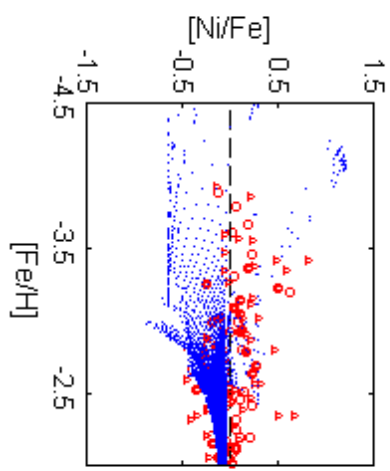
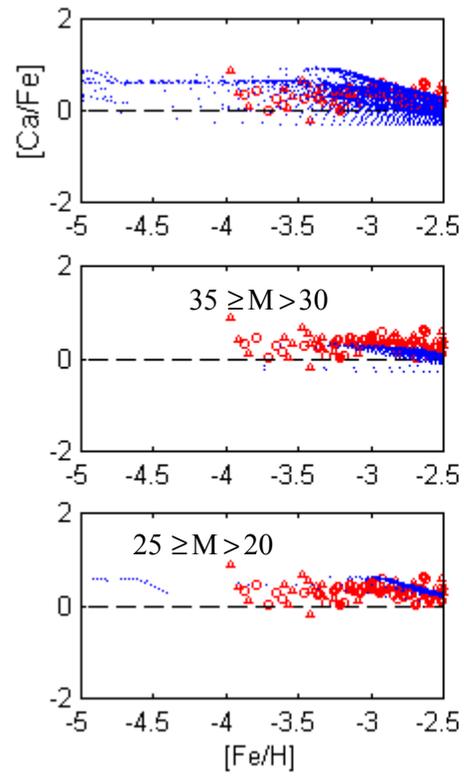
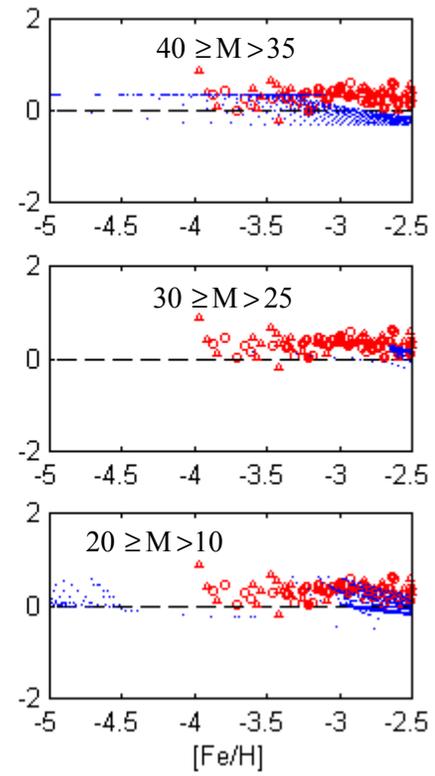
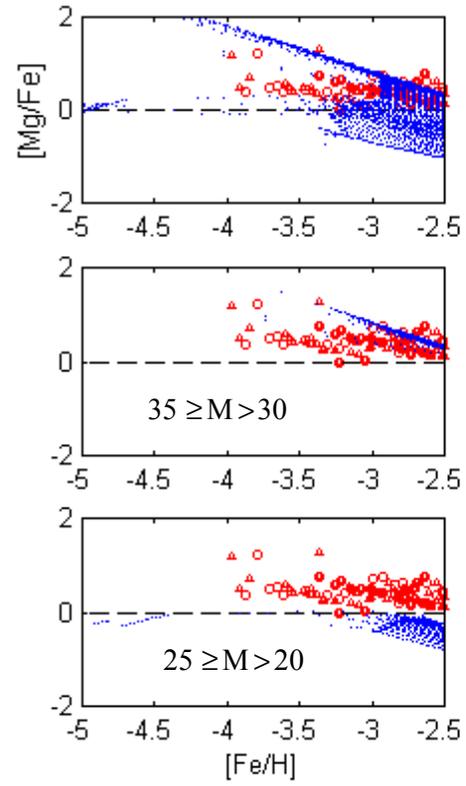
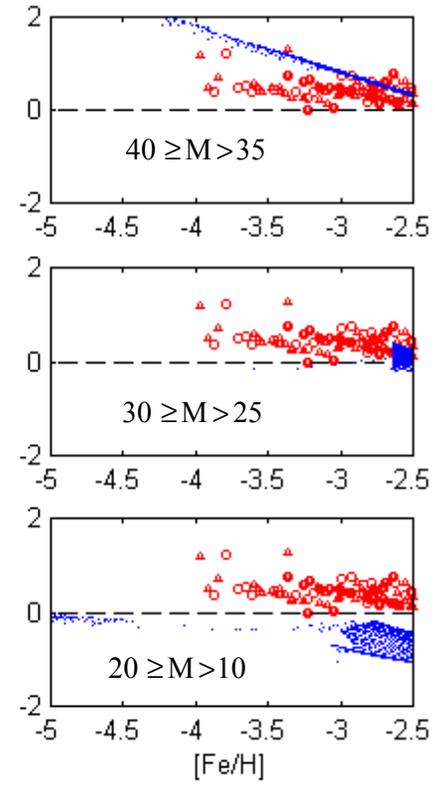
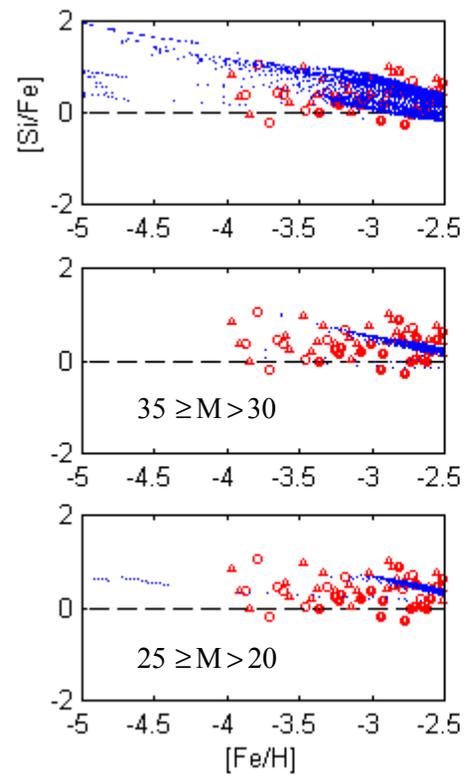
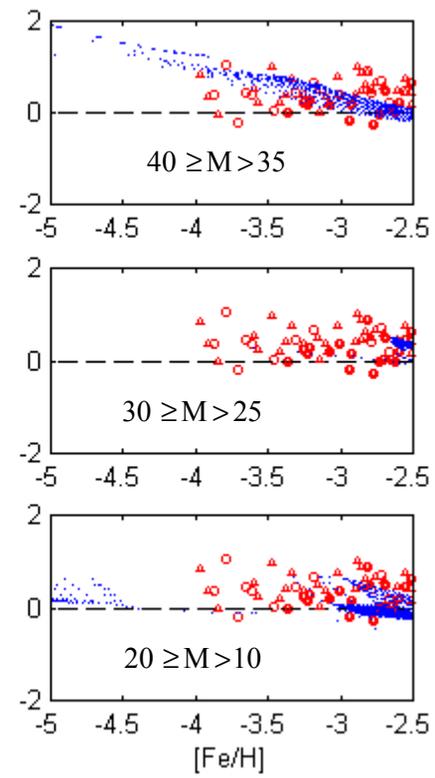
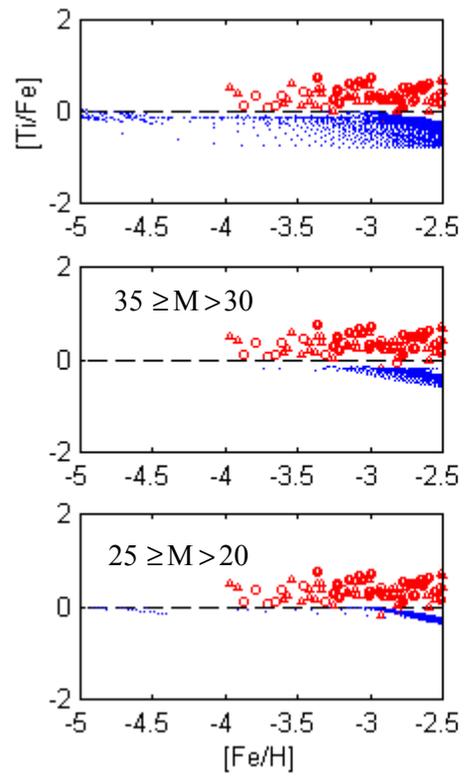
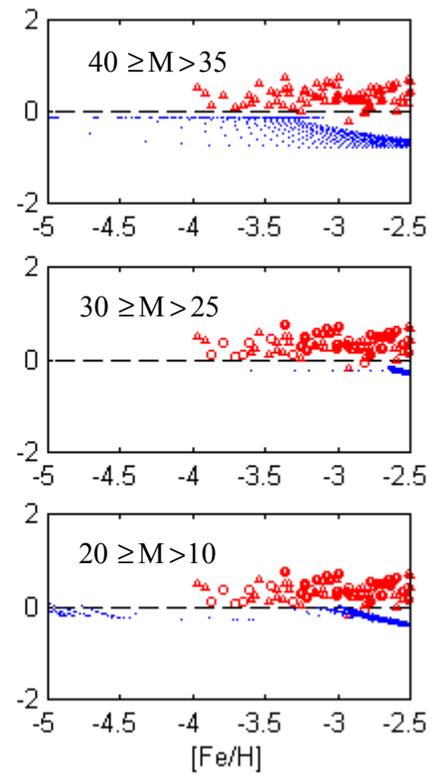
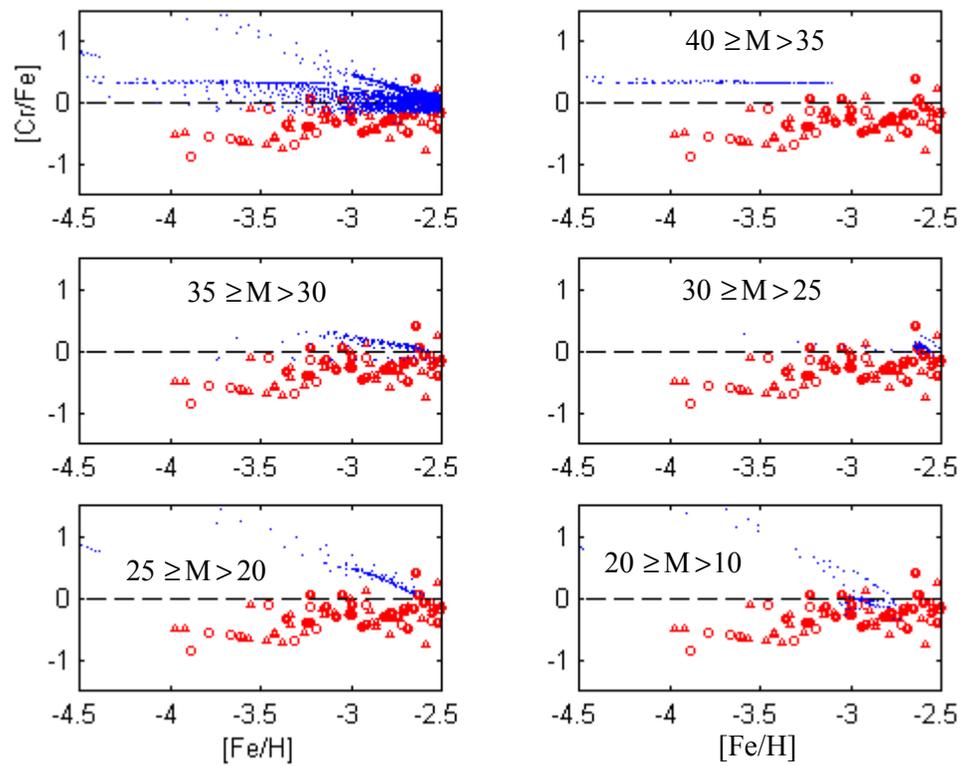
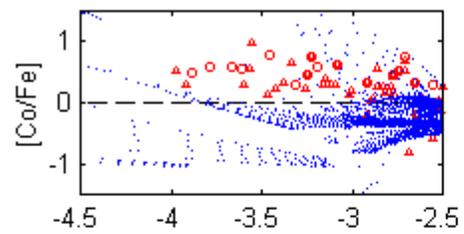
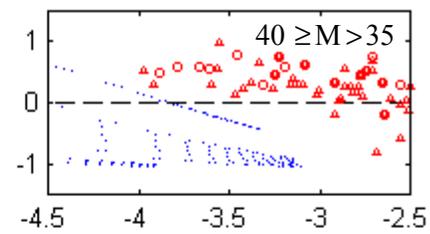
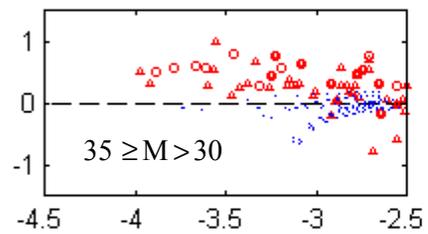
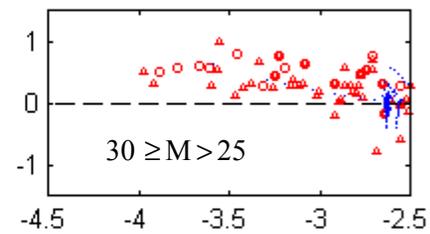
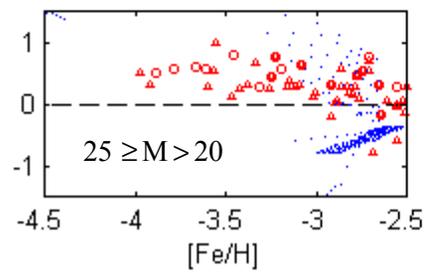
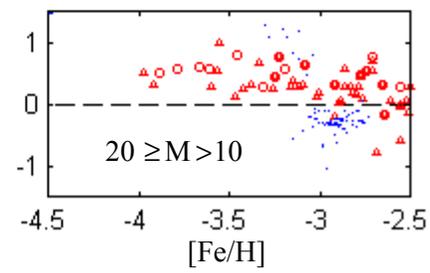
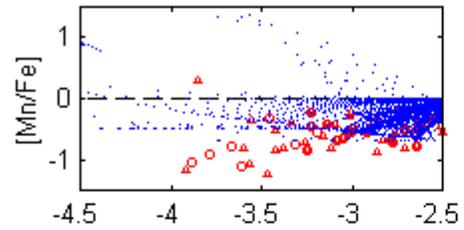
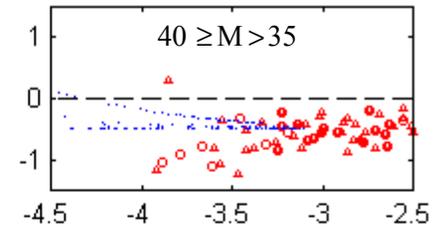
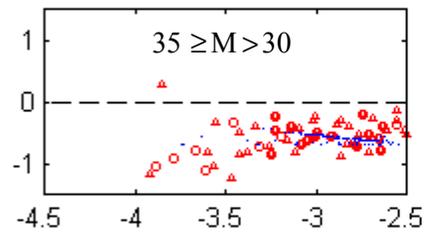
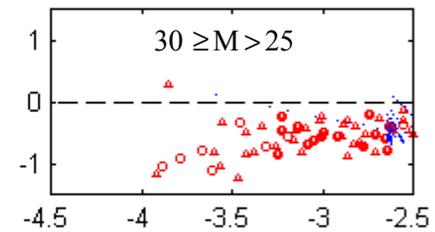
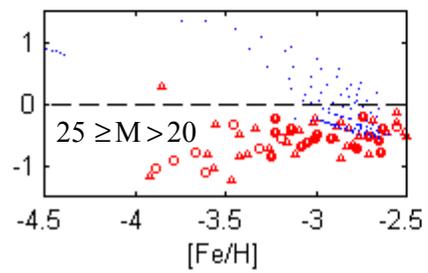
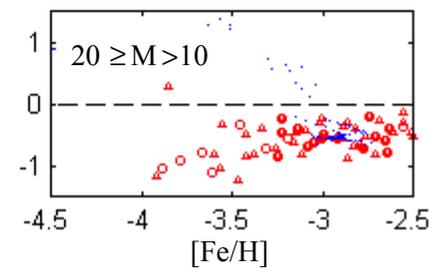
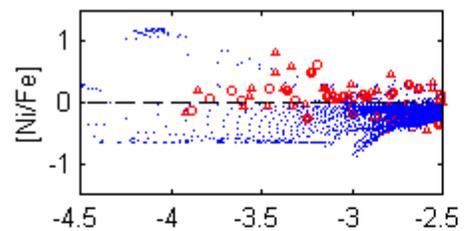
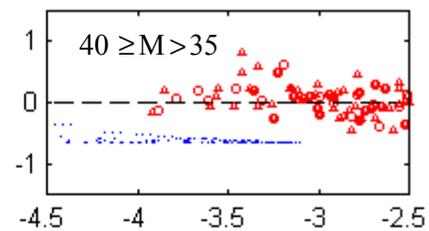
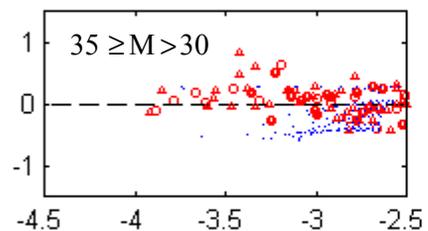
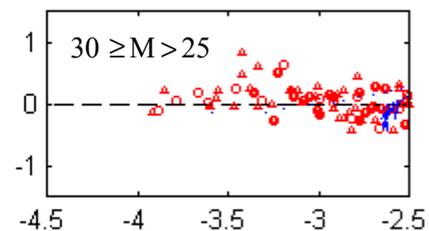
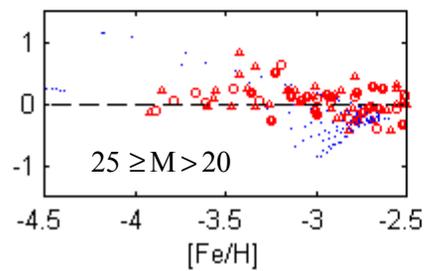
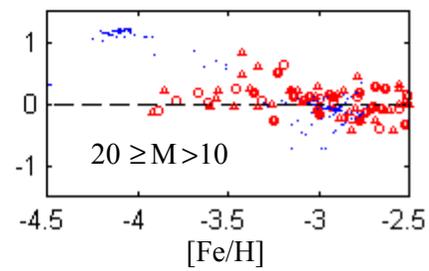
\end{document}